\documentclass[twocolumn, tighten, times, astrosymb, twocolappendix]{aastex631}
\usepackage{amssymb}
\usepackage{amsbsy}
\usepackage{amsfonts}
\usepackage{mathrsfs}
\usepackage{color}
\usepackage{bm}

\defcitealias{Wang2020}{W20}
\defcitealias{Gravity2021b}{G21b}

\def\ihep{State Key Laboratory of Particle Astrophysics, Institute of High Energy Physics, Chinese Academy of Sciences, 19B Yuquan Road, Beijing 100049, China; \href{mailto:liyanrong@ihep.ac.cn}{liyanrong@mail.ihep.ac.cn}, \href{mailto:wangjm@mail.ihep.ac.cn}{wangjm@mail.ihep.ac.cn}}
\def\UCASastro{School of Astronomy and Space Sciences, University of Chinese Academy of Sciences, Beijing 100049, China}
\def\MPE{Max Planck Institute for Extraterrestrial Physics (MPE), Giessenbachstr.1, 85748 Garching, Germany}
\def\KIAA{The Kavli Institute for Astronomy and Astrophysics, Peking University, Beijing 100871, China; \href{mailto:shangguan@pku.edu.cn}{shangguan@pku.edu.cn}}
\def\PKUDoA{Department of Astronomy, School of Physics, Peking University, Beijing 100871, China}
\def\Langrange{Universit\'e Côte d'Azur, Observatoire de la C\^ote d'Azur, CNRS, Laboratoire Lagrange UMR 7293, B\^atiment H. Fizeau, F-06108 Nice Cedex 2, France} 
\def\Johannesburg{Department of Physics, University of Johannesburg, P.O. Box 524, 2006 Auckland Park, Johannesburg, South Africa;}
\def\CAHA{Centro Astronomico Hispano Alem\'an, Sierra de los filabres sn, 04550 gergal. Almer\'ia, Spain}
\def\IAA{Instituto de Astrof\'isica de Andaluc\'ia, Glorieta de la astronom\'ia sn, 18008 Granada, Spain}
\def\NAOC{National Astronomical Observatories of China, Chinese Academy of Sciences, 20A Datun Road, Beijing 100020, China}
\def\YNAO{Yunnan Observatories, Chinese Academy of Sciences, Kunming 650011, China}
\def\DongGuan{Dongguan Neutron Science Center, 1 Zhongziyuan Road, Dongguan 523808, China}
\def\UCASphy{School of Physical Sciences, University of Chinese Academy of Sciences, 19A Yuquan Road, Beijing 100049, China}
\def\WYU{Department of Physics and Astronomy, University of Wyoming, Laramie, WY 82071, USA}

\begin{document}

\title{\bf\large Spectroastrometry and Reverberation Mapping of Active Galactic Nuclei. II. Measuring Geometric Distances and Black Hole Masses of Four Nearby Quasars}

\author[0000-0001-5841-9179]{Yan-Rong Li}
   \affiliation{\ihep}
\author[0000-0002-4569-9009]{Jinyi Shangguan}
   \affiliation{\KIAA}
   \affiliation{\MPE}
\author[0000-0001-9449-9268]{Jian-Min Wang}
   \affiliation{\ihep}
   \affiliation{\UCASastro}
   \affiliation{\NAOC}
\author[0000-0003-4949-7217]{Ric Davies}
\author[0000-0002-5687-0609]{Daryl J. D. Santos}
   \affiliation{\MPE}
\author{Frank Eisenhauer}
   \affiliation{\MPE}
\author[0000-0003-4042-7191]{Yu-Yang Songsheng}
   \affiliation{\ihep}
\author[0000-0003-2662-0526]{Hartmut Winkler}
   \affiliation{\Johannesburg}
\author{Jes\'us Aceituno}
   \affiliation{\CAHA}
   \affiliation{\IAA}
\author{Hua-Rui Bai}
   \affiliation{\ihep}
   \affil{\UCASphy}
\author{Jin-Ming Bai}
   \affiliation{\YNAO}
\author[0000-0002-1207-0909]{Michael S. Brotherton}
   \affiliation{\WYU}
\author[0000-0001-5301-1326]{Yixian Cao}
   \affiliation{\MPE}
\author{Yong-Jie Chen}
   \affiliation{\ihep}
   \affiliation{\DongGuan}
\author[0000-0002-5830-3544]{Pu Du}
   \affiliation{\ihep}
\author{Feng-Na Fang}
   \affiliation{\ihep}
   \affil{\UCASphy}
\author{Jia-Qi Feng}
   \affiliation{\ihep}
   \affil{\UCASphy}
\author{Helmut Feuchtgruber}
   \affiliation{\MPE}
\author[0000-0003-4264-3381]{Natascha M. F\"{o}rster Schreiber}
   \affiliation{\MPE}
\author{Yi-Xin Fu}
   \affiliation{\ihep}
   \affil{\UCASphy}
\author{Reinhard Genzel}
   \affiliation{\MPE}
\author[0000-0002-5708-0481]{Stefan Gillessen}
   \affiliation{\MPE}
\author[0000-0001-6947-5846]{Luis C. Ho}
   \affiliation{\KIAA}
   \affiliation{\PKUDoA}
\author{Chen Hu}
   \affiliation{\ihep}
\author[0000-0003-3086-7804]{Jun-Rong Liu}
   \affiliation{\ihep}
\author[0000-0003-0291-9582]{Dieter Lutz}
   \affiliation{\MPE}
\author{Thomas Ott}
   \affiliation{\MPE}
\author[0000-0003-4759-6051]{Romain G. Petrov}
   \affiliation{\Langrange}
\author{Sebastian Rabien}
   \affiliation{\MPE}
\author[0000-0002-2125-4670]{Taro Shimizu}
   \affiliation{\MPE}
\author[0000-0002-0018-3666]{Eckhard Sturm}
   \affiliation{\MPE}
\author{Linda J. Tacconi}
   \affiliation{\MPE}
%\author{Hannah \"{U}bler}
%   \affiliation{\MPE}
\author{Yi-Lin Wang}
   \affiliation{\ihep}
   \affil{\UCASphy}
\author[0009-0000-1228-2373]{Zhu-Heng Yao}
   \affiliation{\NAOC}
\author{Shuo Zhai}
   \affiliation{\NAOC}
\author[0000-0002-5595-0447]{Hao Zhang}
   \affiliation{\ihep}
   \affil{\UCASphy}
\author{Yi-Peng Zhao}
   \affiliation{\ihep}
   \affil{\UCASphy}
\author[0009-0006-7629-1459]{Yu Zhao}
   \affiliation{\ihep}
   \affil{\UCASphy}
\collaboration{55}{(SARM Collaboration)}

% \correspondingauthor{Yan-Rong Li}
% \email{liyanrong@mail.ihep.ac.cn}

% \correspondingauthor{Jinyi Shangguan}
% \email{shangguan@mpe.mpg.de}

% \correspondingauthor{Jian-Min Wang}
% \email{wangjm@mail.ihep.ac.cn}

\begin{abstract}
The geometric distances of active galactic nuclei (AGNs) are challenging to measure because of their exceptionally compact structure yet vast cosmic distances. A combination of spectroastrometry and reverberation mapping (SARM) of broad-line regions (BLRs) constitutes a novel means to probe the geometric distance of AGNs, which has recently become practically feasible owing to successful interferometric observations with VLTI/GRAVITY. Here, we perform SARM analysis of four nearby quasars: Mrk 509, PDS 456, 3C 273, and NGC 3783. Results for the former two are reported for the first time and the latter two are revisited using our improved BLR dynamical modeling that includes the radial-dependent responsivity of BLRs. This allows us to self-consistently account for the emissivity weighting of the BLR in spectroastrometry and responsivity weighting in reverberation mapping. We obtain angular-diameter distances of the four quasars, from which we derive a Hubble constant of $H_0=69_{-10}^{+12}\,\rm km\,s^{-1}\,Mpc^{-1}$. Although this constitutes a large uncertainty for a measurement of $H_0$, it is anticipated that the precision will improve to a competitive level once a greater number of AGNs are accessible following the upgrade of GRAVITY in the near future. From SARM analysis, the black hole masses of the four quasars are also measured with the statistical uncertainty ranging from 0.06 to 0.23 dex, consistent with the correlations between black hole masses and properties of the host bulges.
\end{abstract}

\section{Introduction}
Since the discovery of the first quasar, hundreds of thousands of active galactic nuclei (AGNs) have been identified from modern large-scale surveys. The popular unification model depicts an overarching generic structure of AGNs (\citealt{Antonucci1993}), but many of the finer details are far from resolved. This is particularly true for the broad-line regions (BLRs) surrounding supermassive black holes (SMBHs), which are responsible for emitting broad emission lines, a definitive characteristic in the broad-band spectra of AGNs. Two observational approaches have been found to be effective and have advanced our understanding of BLR structure and kinematics. The first approach is reverberation mapping  (RM; \citealt{Blandford1982, Peterson1993}) in the time domain, which can be routinely undertaken with common intermediate-aperture telescopes. The other approach probes the spatial domain with the aid of optical interferometry and is referred to as spectroastrometry (SA; \citealt{Beckers1982, Bailey1998, Petrov2001}). It was demonstrated by the successful observation of 3C\,273 \citep{Gravity2018} with the GRAVITY beam combiner mounted on the European Southern Observatory Very Large Telescope Interferometer (VLTI; \citealt{Gravity2017}).

Over the past decades, major efforts have been made to spectroscopically monitor AGNs continually over long periods, enabling measurement of the BLR sizes through RM analysis. The typical sizes of BLRs were found to range from light-days to light-months, mainly depending on the AGN luminosity (e.g., \citealt{Kaspi2000, Bentz2013, Du2019}). Such sizes are too compact to be spatially resolved by modern facilities, even for the nearest AGNs. For instance, the Seyfert 1 galaxy NGC 4151 at a distance of 15.8 Mpc (\citealt{Yuan2020}) has a typical BLR size of $\sim10$ light-days (\citealt{Chen2023}), corresponding to an angular size of $\sim0.1$ milli-arcseconds. For a long time, RM stood out as the most effective method to determine the spatial extent of the BLR (e.g., \citealt{Kaspi2000, Peterson2004, Bentz2009, Du2014}). It has also been an essential tool for the diagnostic of the gas environment in the vicinity of SMBHs (\citealt{Cackett2021}).

Dramatic new opportunities are now becoming available as a result of the GRAVITY instrument coming into operation. GRAVITY combines light from all four 8m Unit Telescopes of the VLTI to form six simultaneous interferometric baselines. It measures the differential phase signal of a broad emission line in the near-infrared $K$ band, which reflects SA of the emission line (namely, photocenter offset as a function of wavelength) with respect to the astrometry of the underlying continuum emission (\citealt{Gravity2018, Li2023}). The corresponding BLR size can then be inferred from the differential phase signal at a level of tens of micro-arcseconds.
Such a high spatial resolution enables resolving the BLR in bright AGNs from the local universe (\citealt{Gravity2018, Gravity2020, Gravity2021a, Gravity2024})
up to $z\sim2$ (\citealt{Gravity+2022, Abuter2024}), provided these have a prominent broad emission line in the $K$ band.

Remarkably, the capabilities of SA and RM techniques are complementary and their combination (hereafter referred to as spectroastrometry and reverberation mapping (SARM)) conveys a more comprehensive view of the BLR geometry and kinematics. This is reflected in the different dimensions probed by SA and RM. To be specific, SA is sensitive to the BLR structure projected onto the observer's sky, i.e., perpendicular to the line of sight. In contrast,  RM is sensitive to the structure perpendicular to the iso-delay surfaces since it measures time delays between variations of the emission line and continuum. The iso-delay surfaces have a paraboloid shape with the focus located at the central ionizing source, typically regarded as a point source coincident with the SMBH. The combination of the results from both techniques thus enables a much better overall understanding of the BLR properties as well as precise SMBH mass measurements.

Another important application of the SARM method is that it directly probes geometric distances to AGNs. The time delays measured from RM represent light travel times from the central ionizing source, typically an accretion disk, to the BLR. Multiplied by the speed of light, these thereby convey the linear size ($\Delta R$) scale of the BLR. The interferometric phases measured from SA reflect the angular astrometry of BLR photons and convey the angular size scale ($\Delta \theta$) of the BLR. A combination of SA and RM naturally constitutes a geometric probe for the angular-diameter distance ($D_{\rm A}=\Delta R/\Delta \theta$) from the basic geometric principles (\citealt{Elvis2002, Rakshit2015, Wang2020, Li2023}). As it is usually difficult to measure both $\Delta R$ and $\Delta \theta$ for the same target, extragalactic objects for which such a geometric method is applicable are exceptionally rare. The SARM approach thus offers a new, promising tool for cosmology, especially given the ubiquitousness and high brightness of AGNs across the Universe.

Hitherto, a sample of seven nearby bright AGNs has been observed by GRAVITY (\citealt{Gravity2018, Gravity2020, Gravity2021a, Gravity2024}). \citet[hereafter \citetalias{Wang2020}]{Wang2020} conducted the first SARM analysis on 3C 273 and \citet[hereafter \citetalias{Gravity2021b}]{Gravity2021b} conducted the second application of the SARM technique on NGC 3783. These initial studies derived an angular-diameter distance $D_{\rm A}=551.5_{-78.7}^{+97.3}$ Mpc with a precision of 15\% for 3C 273, and  $39.9_{-11.9}^{+14.5}$ Mpc with a precision of 30\% for NGC 3783, respectively. They adopted a globally linear, uniform response of the BLR gas. However, there is evidence that the BLR gas might have a radial-dependent, non-linear response.  The most relevant observational phenomenon comes from RM campaigns, which show that the H$\beta$ line widths of the mean spectra are typically broader than those of the root-mean-square (rms) spectra (see Figure 6 in \citealt{LW2024}). This clearly implies that the BLR gas is more responsive at large radial distances, where the BLR gas has a relatively lower velocity (compared to the inner regions), thereby resulting in a narrower line width in the rms spectrum. Such a non-uniform response of BLRs has also been shown from the photoionization calculations (e.g., \citealt{Korista2004, Goad2014, Zhang2021}). If using the terminology ``responsivity'', which quantifies the relative change of the line emissivity with respect to that of the incident continuum flux (e.g., \citealt{Goad1993}), the responsivity of the BLR gas is not constant, but changes with radial distance.  Based on photoionization theory, the main physical reason is as follows. In the inner region where BLR gas receives a higher ionizing photon flux, the excited-state populations of hydrogen and helium atoms are enhanced, leading to a lower efficiency in converting the ionizing photons into recombination lines (such as the Balmer lines) that arise out of these excited states (\citealt{Korista2004}). As a consequence, the BLR gas exhibits a low responsivity. In an extreme case where the BLR gas is fully ionized, the emission lines will no longer respond to any continuum variations, resulting in a zero responsivity. As the distance ($r$) to the ionizing source increases, the incident ionizing photon flux decreases ($\propto r^{-2}$) and the above effect gradually weakens. In other words, for an extended BLR, the changes in the responsivity of emission lines with respect to the incident ionizing photon flux naturally cause a dependence of the responsivity on radial distance.

This radially dependent responsivity of BLR gas has an important implication for SARM analysis. As SA probes emissivity weighting, whereas RM probes responsivity weighting  of the BLR gas, a non-uniform responsivity means that the emissivity- and responsivity-weighted BLR sizes are not the same.
In this work, we perform SARM analysis on two AGNs investigated by \cite{Gravity2024}, Mrk 509 and PDS 456, and revisit the previous analysis on 3C 273 and NGC 3783 using the latest improved BLR dynamical modeling that includes radially dependent responsivity of the BLR gas (\citealt{LW2024}). This allows us to self-consistently deal with emissivity weighting in SA and responsivity weighting in RM and therefore to alleviate inaccuracies arising from the adoption of globally uniform responses in previous analysis.

The paper is organized as follows. Section~\ref{sec_data} compiles the observation data of SA from VLTI/GRAVITY and RM from previous optical spectral monitoring campaigns. Section~\ref{sec_sarm} introduces the methodology of SARM joint analysis, including continuum light curve modeling, BLR dynamical modeling, and Bayesian inference. Section~\ref{sec_results} summarizes estimates of the geometric distances, Hubble constant, and the black hole masses.
Section~\ref{sec_dis} presents discussions on possible systematic uncertainties in the determined distances, followed by the conclusions in Section~\ref{sec_conclusion}. Appendix~\ref{app_size} defines three types of BLR sizes, Appendix~\ref{app_parameters} lists the full set of model parameters and their inferred values, Appendix~\ref{app_sa} shows the best fits to the differential phases of the three baselines with the strongest BLR signals, Appendix~\ref{app_spec} plots the mean and rms spectra of the four AGNs, Appendix~\ref{app_etamax} tests the influence of the upper limit of responsivity on the inferred model parameters, and Append~\ref{app_rm} compares the obtained time lags from BLR dynamical modeling with those derived from the traditional cross-correlation analysis of the RM data.

\begin{deluxetable*}{ccccc}
\tablecaption{Objects and the Observation Periods of GRAVITY and RM data.\label{table_object}}
\tablehead{
\colhead{}  & \colhead{Mrk 509}  & \colhead{PDS 456} &  \colhead{3C 273}  & \colhead{NGC 3783}
}
\startdata
$z$ & 0.0344 & 0.185  & 0.15834 & 0.009730\\
GRAVITY & Jul. 2021 (2 nights) &  Aug. 2018-Jul. 2021 (9 nights) & Jul. 2017-May 2018 (8 nights) & Jan. 2018-Mar. 2020 (6 nights)\\
RM & May 2019-Nov. 2022  &  May 2020-Oct. 2023            & Nov. 2008-Jul. 2018           & Feb. 2020-Jun. 2020\\
\enddata
\end{deluxetable*}

\section{Observation Data}\label{sec_data}
We collected the GRAVITY interferometric data and RM data of four AGNs: Mrk 509, PDS 456, 3C 273, and NGC~3783.
Table~\ref{table_object} summarizes the observation periods (i.e., time range of the data) and the redshift used in our analysis. Below we briefly describe the main observational information as well as our additional reduction procedure to the GRAVITY interferometric data.

\subsection{GRAVITY Interferometric Data}
Mrk 509 was observed with GRAVITY on 25-26 July, 2021 with total on-source exposure time of 100 minutes.
PDS~456 was observed during several epochs between 2018 and 2021, with total on-source exposure time of 295 minutes.
The interferometric data of these two AGNs were reported
in \cite{Gravity2024}, together with the results for two other AGNs, Mrk 1239 and IC 4329A. As these latter two are subject to large noise and the differential phase signals are less prominent, we only focus on Mrk~509 and PDS~456 in the present work.
3C 273 was observed with GRAVITY on eight nights between July 2017 and May 2018 and the interferometric data was reported in \cite{Gravity2018}.
NGC~3783 was observed on 6 nights between January 2018 and March 2020 and the interferometric data was reported in \cite{Gravity2021a}.

We made additional reductions to the differential phase and line profile data of Mrk~509 and
PDS~456 after applying the pipeline following the procedures described
in \cite{Gravity2020}.  We first fit and remove the static and variable 
instrumental effect of the differential phase for individual exposures.  
We also use a first-order polynomial function to remove the residual phase 
at the continuum emission wavelength (see Section~2.3 of 
\citealt{Gravity2020}).  We determine the phase error from the standard 
deviation of the phase at the continuum for each exposure and average all 
the exposures weighted according to their phase errors.  To calibrate the broad 
Br$\gamma$ or Pa$\alpha$ line, we used the calibrator stars observed before or 
after the AGN observations to remove the instrumental and telluric features.  
Thereafter, we apply a third-order polynomial function to fit the calibrated continuum and
normalize the line profile to the continuum.  For Mrk~509 and PDS~456, we prefer 
using early-type stars to calibrate the spectra because they have less
absorption features, with only the Br$\gamma$ absorption line significant in the relevant spectral range, which imprints a narrow Br$\gamma$ emission line on the final calibrated AGN spectra.  To subtract this contaminating component, we use two Gaussian functions to fit the AGN Br$\gamma$ or Pa$\alpha$ line and a Lorentzian function to fit
the narrow Br$\gamma$ line.  
The AGN and stellar Br$\gamma$ lines are not strongly blended (if at all) because 
the AGN Br$\gamma$ line is significantly redshifted.  We notice that, compared to the double-Gaussian function used to fit the stellar Br$\gamma$ line as by 
\cite{Gravity2024}, the Lorentzian function can improve
the fit to the Pa$\alpha$ line profile of PDS~456.  This is because the double-Gaussian function for the stellar Br$\gamma$ line cannot be properly separated from those for the AGN Pa$\alpha$ line, thereby causing an inappropriate subtraction around the blue wing of the AGN line.  We directly adopt the data of 3C~273 and NGC~3783
from \cite{Gravity2018} and \cite{Gravity2021a}, respectively.  The NGC~3783 
data were reduced in the same way as described above.
The 3C~273 data were, however, reduced differently.  Instead of
decomposing the instrumental differential phase signatures into static and 
variable components, \cite{Gravity2018} incorporated a Gaussian kernel with 
an FWHM of 24 pixels to convolve the differential phase spectra and generate the broad instrumental differential phase.  We confirmed that it makes little difference if we conduct the joint analysis with the data reduced with the new approach.

%{\bf a brief mention of improved data reduction if necessary...to be complemented by Shangguan...}

\subsection{Reverberation Mapping Data}\label{sec_rmdata}
\cite{Li2024} reported multi-year RM data of five AGNs (IC 4329A, Mrk 335, Mrk 509, Mrk 1239, and PDS 456), selected from the sample of potential AGNs suited for GRAVITY/GRAVITY+ observations compiled by \citetalias{Wang2020}. For Mrk~509 and PDS 456, the optical spectroscopic and photometric monitoring were conducted from May 2019 to November 2022, and from May 2020 to October 2023, respectively, with a typical sampling cadence of every 4-5 days. The photometric data were further combined with the public archival data from the Zwicky Transient Facility (ZTF; \citealt{Graham2019}) and the All-Sky Automated Survey for Supernovae (ASAS-SN; \citealt{Shappee2014, Kochanek2017}) to yield high-cadence continuum light curves. These different sources of data were
intercalibrated with the Bayesian software PyCALI\footnote{The latest code is publicly available at \url{https://github.com/LiyrAstroph/
PyCALI}, while the version used in this work is available in \cite{pycali}.} (\citealt{Li2014}), which uses a multiplicative factor and an additive factor to account for the inhomogeneous apertures and differences in flux calibration. For Mrk~509, from the four years of monitoring, the H$\beta$ time delay (around 50 days) can be detected in three years (2019, 2021, and 2022). However, when using the full data span, there appears to be a long-term trend in the H$\beta$ variations, leading to an apparently long time delay component (around 300 days). The explanation for such a long-term trend is still under debate. We therefore only use one-year data for SARM analysis. We chose the RM data for the year 2022 since it shows the most significant variability.  We also present analysis with the RM data in 2019 and 2021 and find that the obtained distances are consistent with each other within uncertainties. For PDS 456, the H$\beta$ time delay is about 250 days and we use the four-year RM data for the SARM analysis.

For 3C 273, \cite{Zhang2019} presented the analysis of 10 years of optical monitoring data between November 2008 and July 2018, using the public archival data from the Steward Observatory spectropolarimetric monitoring project (\citealt{Smith2009}) and new observations from the super-Eddington accreting massive black hole project (e.g., \citealt{Du2014}). 3C 273 is radio loud and the optical emission from the beamed jet is not negligible compared to that from the accretion disk. As demonstrated by \cite{Li2020}, the jet contamination results in a distinct long-term component in the continuum light curve that does not have a corresponding echo in the H$\beta$ fluxes. Such jet contamination can be partly corrected by subtracting a linear long-term trend. As we show below, we set the slope of the linear trend as a free parameter to be determined from the Bayesian inference.

The latest published RM study of NGC 3783 was conducted by \cite{Bentz2021} throughout the ﬁrst half of 2020 using the Las
Cumbres Observatory global telescope network. A total of 50 spectra were obtained with a typical sampling interval of 2 days.
The 5100~{\AA} data was merged with the $V$-band photometric data (with 209 epochs), resulting in a continuum light curve
spanning a time span of 137 days from JD 2,458,891 to 2,459,028 (see Figure 4 in \citealt{Bentz2021}). NGC 3783 has also been covered by the ASAS-SN survey since February 2012. We use the PyCALI software to scale and shift the ASAS-SN light curve to match the merged light curve of \cite{Bentz2021}.
This extends the temporal baseline of the continuum light curve from 137 days to 268 days in the 2020 season and also supplies continuum variation information for other seasons. The former is useful to constrain the BLR model, and the latter helps to better characterize the continuum variability.

In modeling RM data, we need to subtract the components in the continuum light curves irrelevant to  emission-line reverberation. One such component is the host galaxy's starlight. Among the four AGNs, Mrk 509, PDS~456, and 3C~273 have a high optical luminosity at 5100~{\AA} $L_{\rm 5100}>10^{44.3}~{\rm erg~s^{-1}}$ (\citealt{Zhang2019, Li2024}) and the host galaxy contribution can be treated as negligible. This is supported by the spectral decomposition, which did not identify a significant host galaxy component (\citealt{Zhang2019, Li2024}). For NGC~3783, we subtract the host galaxy component ($f_{\rm gal}=2.76\times10^{-15}~\rm erg\,s^{-1}\,cm^{-2}\,\text{\AA}^{-1}$) according to the Hubble Space Telescope image decomposition employed by \cite{Bentz2021}. 3C~273 is additionally subject to the contamination from the jet emission. We subtract an averaged jet flux from the continuum light curve according to the decomposition of the jet and accretion disk emissions proposed by \cite{Li2020}, who derived an averaged jet flux to the total continuum flux fraction of $\sim0.28$.

As summarized in Table~\ref{table_object}, we note that  the GRAVITY and RM observations of all the four objects are undertaken in overlapping periods. This largely minimizes the influences of possible secular changes in BLR kinematics, which occur in the dynamical timescale, typically of the order of several years or longer.

\section{SARM Joint Analysis}\label{sec_sarm}
The procedure underpinning the SARM joint analysis has been described in a number of previous studies, such as \citetalias{Wang2020}, \citetalias{Gravity2021b}, \cite{Li2022}, and \cite{Li2024}. What distinguishes our analysis from these earlier studies is that
we include the latest improvement of BLR dynamical modeling that takes into account radial-dependent responsivity of the BLR (see \citealt{LW2024} for details). This improvement allows us to accommodate for responsivity weighting in SA and emissivity weighting in RM (e.g., \citealt{Zhang2021, Li2022, LW2024}). For the sake of completeness, we briefly describe the major points of the SARM analysis below.

To correct for the cosmological time dilation, we reduce the observation times of the RM light curves by a factor of $(1+z)$, where $z$ is the redshift of the object. Since the RM analysis depends only on the time differences between data points, this manipulation equivalently converts the light curves into the rest frame.

\subsection{Continuum Light Curve Modeling}
We first need to evaluate the continuum fluxes on a sufficiently well-sampled time grid, which is required to integrate reprocessed emission from the whole BLR with a wide range of time delays. To this end, we employ the damped random walk (DRW) model to delineate the continuum variability, which is a stationary Gaussian process with well-established procedures to interpolate and extrapolate the observed continuum light curve (e.g., \citealt{Rybicki1992, Foreman2017, Li2018}). We use a series of evenly spaced points to represent the continuum reconstruction. This series needs to cover an adequate time period prior to the first epoch of the emission line so as to appropriately model the emission-line flux at any given observation epoch. Specifically, the time length of this period should be equal to or longer than the maximum time delay of the BLR. These continuum points plus two hyperparameters of the DRW process (the damping time scale and variation amplitude) are treated as free parameters, and their best-fit values are determined during the subsequent Bayesian inference (see below).

\subsection{BLR Dynamical Modeling}\label{sec_model}
We construct a BLR dynamical model following \cite{LW2024}, which made improvements to the previous modeling scheme of \cite{Pancoast2014} and \cite{Li2018} by including radial-dependent responses. The BLR is assumed to be composed of a large number of point-like, non-interacting clouds, moving in the gravitational well of the central SMBH. These clouds reprocess the ionizing continuum emissions into line emissions, with variable emissivity and responsivity depending on the clouds' radial location. Here, responsivity quantifies the relative change of line emissivity with respect to that of the incident continuum\footnote{Responsivity is mathematically defined as $\eta=d\ln \epsilon/d\ln f_c$, where $\epsilon$ is the cloud's emissivity and $f_c$ is the ionizing flux illuminating the cloud.}  (\citealt{Korista2004}).
The optical continuum is used as a surrogate to the unobservable ionizing continuum as their inter-band time delay is negligible compared to the typical time delays of the BLR.

BLR clouds are distributed within a thick disk, with an opening angle $\theta_{\rm opn}$ and inclination angle $\theta_{\rm inc}$ (see Figure~1 of \citealt{Li2013}).
The radial distribution of the clouds follows a shifted gamma distribution, namely, the cloud's radial distance from the central black hole is given by $r=R_{\rm S} + F R_{\rm BLR} + \beta^2 R_{\rm BLR}(1-F)g$, where $R_{\rm S}$ is the Schwarzschild radius, $R_{\rm BLR}$ is the mean BLR radius, $F$ is the fraction of the inner BLR edge relative to $R_{\rm BLR}$, and $g$ is a random number drawn from a gamma distribution $p(x|a, b)\propto x^{a-1}e^{-x/b}$ with $a=\beta^{-2}$ and $b=1$. The parameter $\beta$ quantifies the standard deviation of the BLR cloud's radial distance from the central point in units of $R_{\rm BLR}$. It also controls the shape of the radial distribution, which may be made to be exponential ($\beta=1$),
long-tailed ($\beta>1$), or narrow peaked ($\beta<1$). Within the opening angle, BLR clouds are distributed uniformly in the azimuthal $\varphi$-direction. In the vertical $\theta$-direction, the angular displacement of a cloud is assigned as $\theta = \cos^{-1}[\cos\theta_{\rm opn}+(1-\cos\theta_{\rm opn})\times u^{\gamma}]$, where the parameter $\gamma$ controls the extent to which clouds are clustered over the outer face of the BLR disk and $u$ is a random number drawn from a uniform distribution between 0 and 1 (\citealt{Pancoast2014}).

The motion of BLR clouds is parameterized to account for various kinematics, including elliptical Keplerian orbiting and bound and unbound inflow/outflow (\citealt{Pancoast2014}). A fraction $f_{\rm ellip}$ of clouds are in elliptical Keplerian orbits and the remaining fraction $(1-f_{\rm ellip})$ of clouds are either inflowing or outflowing. The latter is controlled by a parameter $f_{\rm flow}$, with $f_{\rm flow}\leq0.5$ indicating inflow and $f_{\rm flow}>0.5$ indicating outflow. The velocity of each cloud is first determined for the orbital plane in which it moves around the central SMBH and then converted into real three-dimensional velocity through coordinate transformation. For elliptical orbits, radial and tangential velocities in the $v_r-v_\varphi$ plane are drawn from Gaussian distributions centered around the point $(0, v_{\rm circ})$ along the ellipse connecting circular orbiting velocity ($v_{\rm circ}=\sqrt{GM_\bullet/r}$) and radial escape velocity ($=\sqrt{2}v_{\rm circ}$). For inflowing/outflowing clouds, velocities are assigned similarly, except that the Gaussians are centered around
$(\pm \sqrt{2}v_{\rm circ}, 0)$. In addition, the Gaussian centers are allowed to shift along the ellipse, controlled by a parameter $\theta_e$, to mimic the possibility that clouds are inflowing or outflowing at a velocity smaller than the escape velocity (see \citealt{Pancoast2014}). It is important to note that the SA signal is sensitive to the degree of coherent rotation of the BLR clouds. All clouds' tangential velocities ($v_\phi$) are forced to have the same sign (e.g., positive) to ensure they rotate coherently.

From photoionization calculations (\citealt{Korista2004}), it is known that the reprocessing coefficient of BLR clouds, denoted by $A$,  has a radial dependence.
The reprocessing coefficient $A$, together with the radial distribution of clouds, determines the broad-line emissions.
%However, as a phenomenological model, $A$ is completely degenerate with the radial distribution of clouds.
To reduce the degrees of freedom, we let $A$ be absorbed into the parameterization of the radial distribution and simply set $A=1$ for all clouds. This means that the radial distribution of BLR clouds includes emissivity weights and should not simply be treated as the number density distribution.

We now introduce a parameterization of the responsivity of clouds as (\citealt{LW2024})
\begin{equation}
\eta(r)=\eta_0 + \eta_1\left(\frac{r}{R_{\rm BLR}}\right)^\alpha,
\label{eqn_eta}
\end{equation}
where $\eta_0$, $\eta_1$, and $\alpha$ are free parameters. This form of responsivity can approximately reproduce the photoionization calculations in which the BLR clouds are more responsive at large radius (e.g., \citealt{Korista2004, LW2024}). In our calculations, the responsivity is further limited to a range $0\leq\eta(r)\leq1.5$ and those $\eta$ outside this range are forced to take the lower/upper limits. The lower limit is set to ensure positive responses of BLR clouds to the continuum variations, while the upper limit is set to align with previous photoionization calculations, which generally found $\eta<1.5$ (e.g., \citealt{Korista2004, Goad2014, Goad2015, Zhang2021, LW2024}).
We also include the following two anisotropic effects. Each cloud is assigned an emission weight $w=1/2 + \kappa\cos\phi$, where $\kappa$ is a free parameter and $\phi$ is the angle between the observer's and cloud's line of sight to the central continuum source. In addition, the transparency of the BLR equatorial plane is controlled by a parameter $\xi$, which represents the fraction of clouds below the equatorial plane that is not obscured and contributes to the observed line emissions.

With the application of the above BLR model, the emissivity- and responsivity-weighted transfer functions at time delay $\tau$ and velocity $v$ are obtained by summing over all clouds (\citealt{LW2024})
\begin{eqnarray}
\Psi_e(\tau, v) &=& \sum_i w_i \delta(\tau-\tau_i)\delta(v-v_i),\label{eqn_psie}\\
\Psi_r(\tau, v) &=& \sum_i \eta_i w_i \delta(\tau-\tau_i)\delta(v-v_i),\label{eqn_psir}
\end{eqnarray}
where $\delta(x)$ is the Dirac delta function and the subscript $i$ refers to the $i$-th cloud.

\subsection{Calculating Flux, Profile, and Spectroastrometry of the Emission Line}
By recasting the continuum into constant and variable components $F_c(t) = \bar F_c+\Delta F_c(t)$, the emission line flux is calculated as (\citealt{LW2024})
\begin{equation}
F_l(t) = \bar F_c\int\Psi_e(\tau)d\tau  + \int \Delta F_c(t-\tau)\Psi_r(\tau) d\tau,
\label{eqn_fl}
\end{equation}
where $\Psi_e(\tau)$ and $\Psi_r(\tau)$ are velocity integrals of the transfer functions calculated by Equations~(\ref{eqn_psie}) and (\ref{eqn_psir}).
On the right-hand side of Equation~(\ref{eqn_fl}), the first term represents the constant component and the second represents the variable component of the emission line light curve. The emission-line profile is given by
\begin{equation}
F_l(v) = \bar F_c \Psi_e(v),
\end{equation}
where $\Psi_e(v)$ is the delay integral of the emissivity-weighted transfer function.

The differential phase at velocity $v$ is calculated as
\begin{equation}
\Delta\phi(v) = -2\pi \frac{f_l(v)}{1+f_l(v)}\frac{\bm{u}\cdot(\bm{x}_{\rm BLR}-\bm{x}_c)}{D_{\rm A}},
\end{equation}
where $f_l(v)$ is the flux ratio of the emission line relative to the underlying continuum, $\bm{u}$ is the $uv$ coordinate of the baseline, $\bm{x}_{\rm BLR}$ is the emissivity-weighted offset of the BLR projected onto the sky, $\bm{x}_c$ is the offset of the continuum emission, and $D_{\rm A}$ is the angular-diameter distance. The offset $\bm{x}_{\rm BLR}$ is determined by summing over all clouds as
\begin{equation}
\bm{x}_{\rm BLR}(v) = \sum_i w_i \bm{x}_i \delta(v-v_i),
\label{eqn_offset}
\end{equation}
where $\bm{x}_i=\bm{r}_i - \bm{r}_i\cdot\bm{n}_{\rm los}$ is the $i$-th cloud's offset, $\bm{r}_i$ is the $i$-th cloud's three-dimensional coordinate, and  $\bm{n}_{\rm los}$ is the unit vector in the line of sight.

It is worth reiterating that because of the radial-dependent responsivity, the emissivity- and responsivity-weighted BLR sizes are no longer the same (see Appendix~\ref{app_size} for the definitions). For a radially increasing responsivity ($\eta_1>0$ and $\alpha>0$), the BLR clouds are more responsive at larger radius. This causes the responsivity-weighted size obtained from RM to be larger than the emissivity-weighted size obtained from SA (\citealt{LW2024}). The inclusion of responsivity in dynamical modeling enables us to self-consistently combine these two observational techniques and eliminate any potential biases that may arise from imposing a uniform BLR response in SARM analysis.

%=============================================================================================
\begin{figure*}[th!]
\centering
\includegraphics[width=0.98\textwidth]{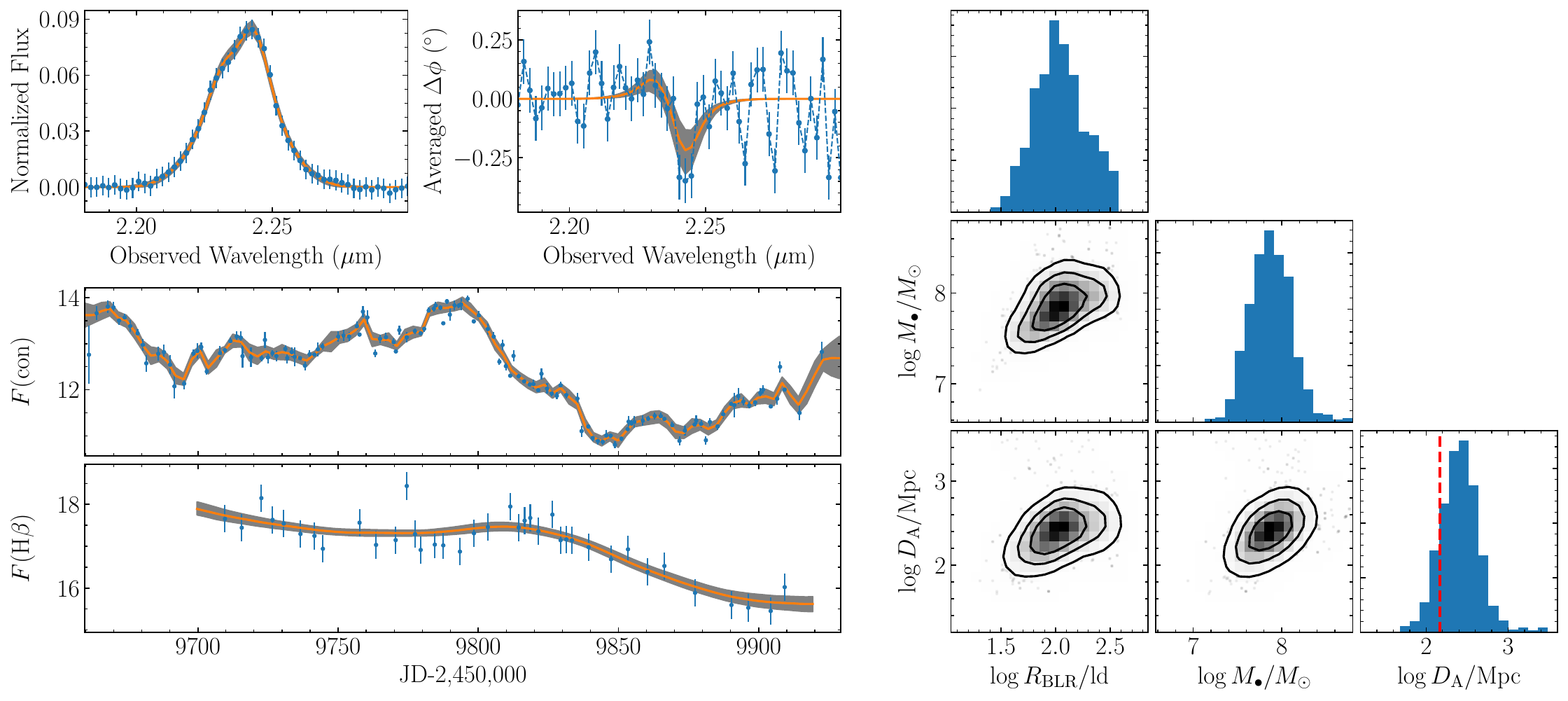}
\caption{SARM analysis results of Mrk 509. The two upper left panels show the normalized profile and averaged differential phase of the Pa$\alpha$ line from the three baselines with the strongest BLR phase signals (see Appendix~\ref{app_sa}). The continuum phase has been subtracted from the differential phases. The two bottom left panels show the continuum and H$\beta$ light curves. Solid orange lines with grey shaded bands (68.3\% confidence interval) show the best reconstructions from model fits. The continuum flux density is in units of $10^{-15}~{\rm erg\,s^{-1}\,cm^{-2}\,\text{\AA}^{-1}}$ and the H$\beta$ flux is in units of $10^{-13}~{\rm erg\,s^{-1}\,cm^{-2}}$. The right panels show posterior distributions of the BLR size ($R_{\rm BLR}$), black hole mass ($M_\bullet$), and angular-diameter distance ($D_{\rm A}$). The contours are at the 1$\sigma$, 1.5$\sigma$, and 2$\sigma$ levels. In the panel for the distribution of $D_{\rm A}$, the red dashed vertical line represents the predicted distance from a flat $\Lambda$CDM cosmology with $H_0$=67 km~s$^{-1}$~Mpc$^{-1}$ and $\Omega_{\rm M}=0.3$.}
\label{fig_Mrk509}
\end{figure*}

\begin{figure*}[th!]
\centering
\includegraphics[width=0.98\textwidth]{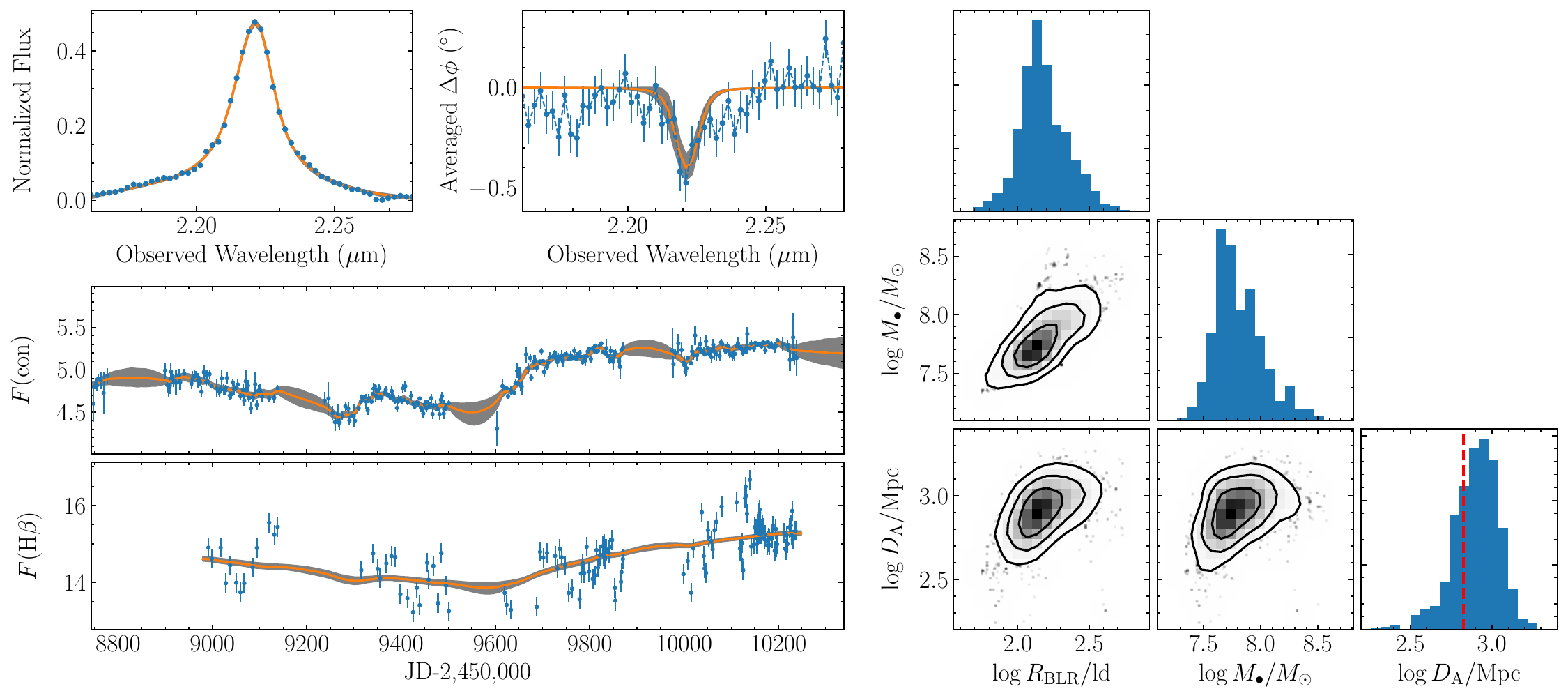}
\caption{Same as Figure~\ref{fig_Mrk509}, but for PDS 456. }
\label{fig_PDS456}
\end{figure*}
%=============================================================================================

%=============================================================================================
\begin{figure*}[ht!]
\centering
\includegraphics[width=0.98\textwidth]{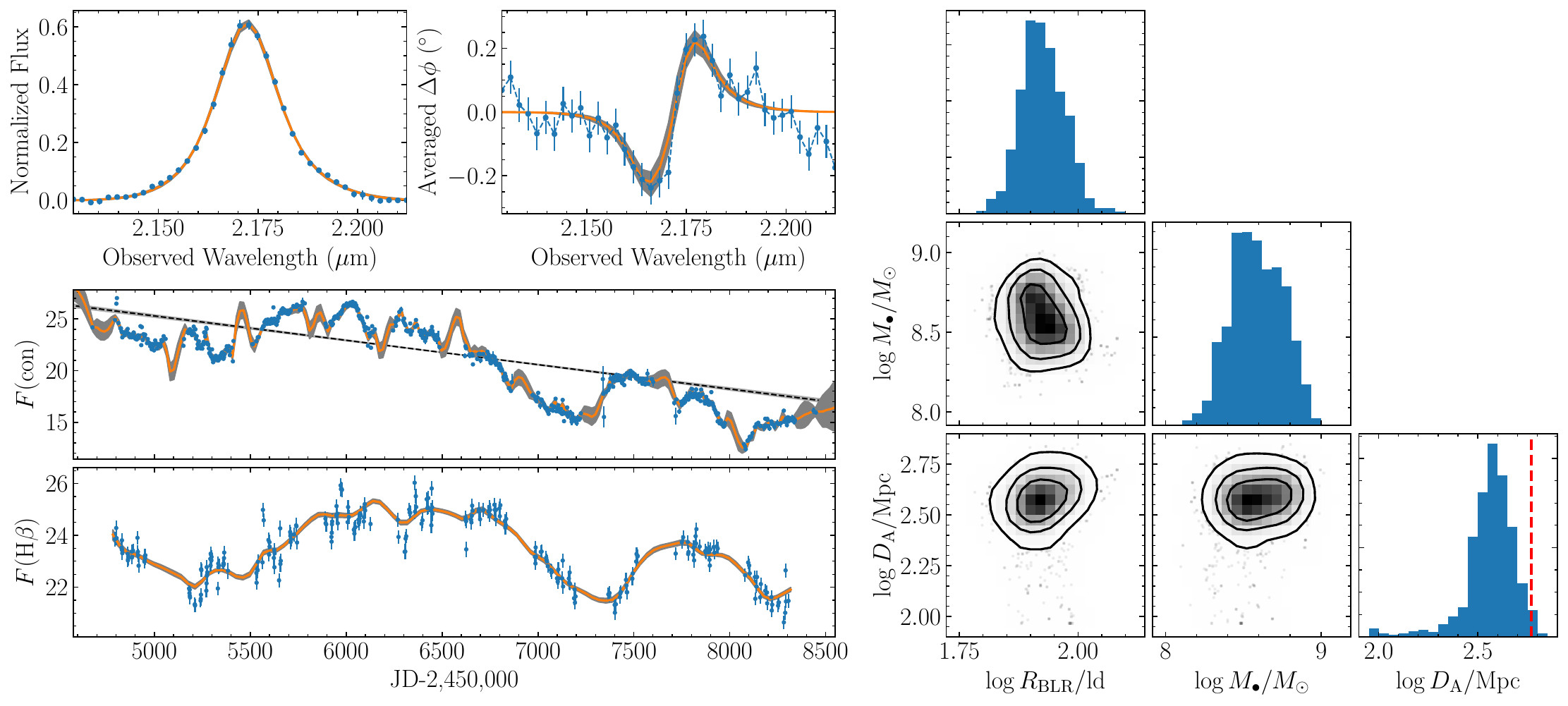}
\caption{Same as Figure~\ref{fig_Mrk509}, but for 3C 273. In the left middle panel, the dashed line represents the linear polynomial used to detrend the continuum light curve.  }
\label{fig_3C273}
\end{figure*}
\begin{figure*}[ht!]
\centering
\includegraphics[width=0.98\textwidth]{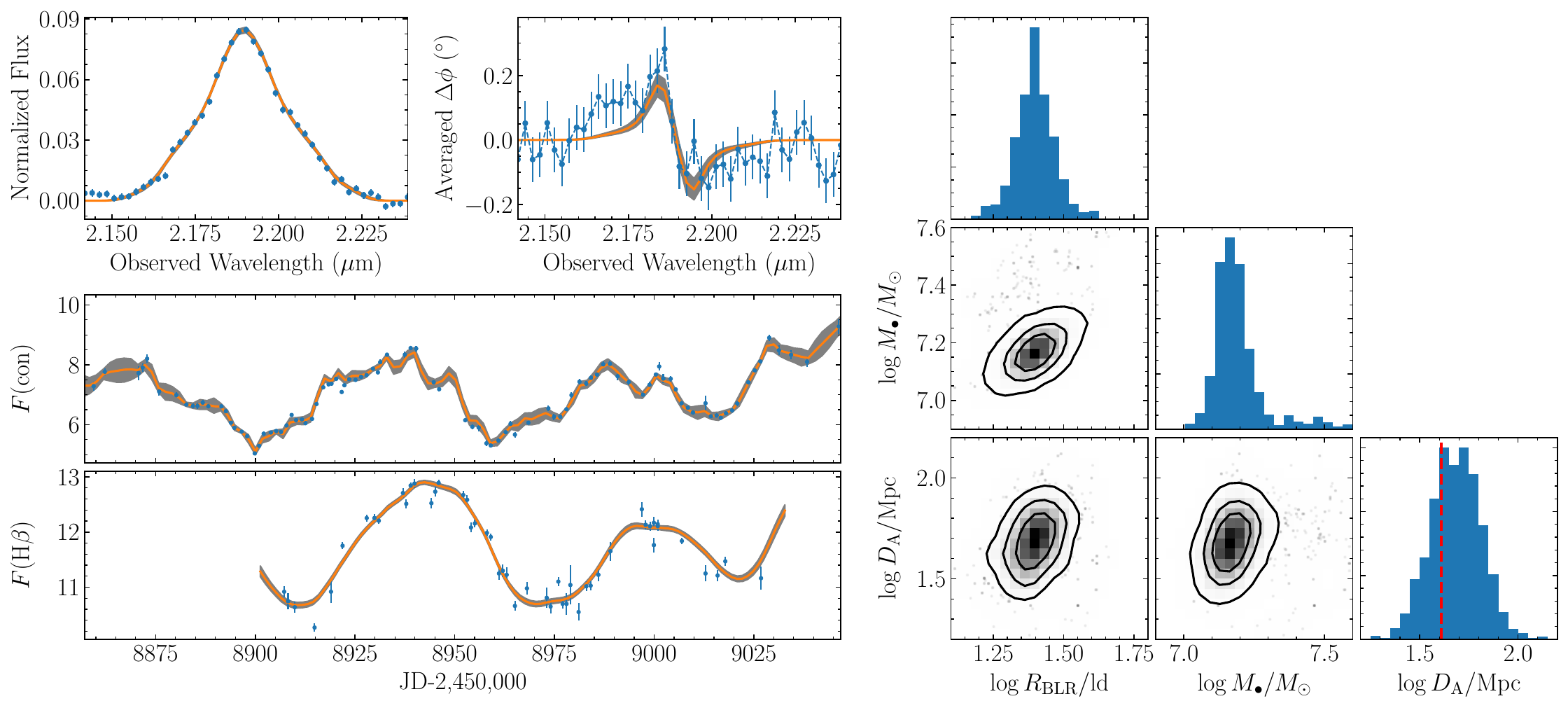}
\caption{Same as Figure~\ref{fig_Mrk509}, but for NGC 3783. }
\label{fig_NGC3783}
\end{figure*}

\begin{figure*}[ht!]
\centering
\includegraphics[width=0.95\textwidth]{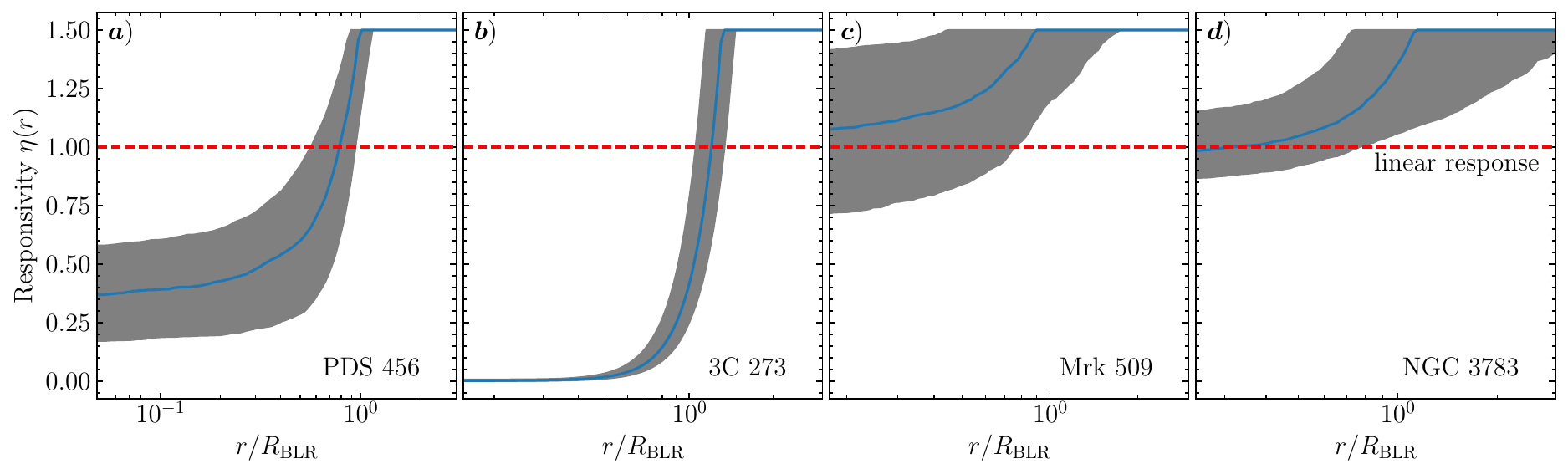}
\caption{The inferred responsivity ($\eta$) as a function of radial distance for the four quasars, a) PDS 456, b) 3C 273, c) Mrk 509, and d) NGC 3783. The form of $\eta$ is given by Equation~(\ref{eqn_eta}). The solid blue lines represent the median values and the grey shaded bands represent the 68.3\% confidence intervals calculated from the posterior samples. The maximum values of $\eta$ is limited to 1.5 to align with photoionization calculations. The horizontal red dashed line represents the linear response ($\eta=1$).}
\label{fig_eta}
\end{figure*}
%=============================================================================================

\subsection{Bayesian Inference}
Given the above continuum light curve and BLR dynamical modeling, we can calculate the modeled SA and RM data and compare these against observations to determine the posterior distributions of model parameters using the Markov Chain Monte Carlo (MCMC) technique.
The SA data include the profile and differential phases of the broad Pa$\alpha$ or Br$\gamma$ line.
The corresponding likelihood function is written as
\begin{eqnarray}
\mathcal{L}_{\rm SA} &=& \prod_i \frac{1}{\sqrt{2\pi}\sigma_{\phi, i}}\exp\left\{ - \frac{[\Delta\phi_i-\Delta\phi_i^m]^2}{2\sigma_{\phi, i}^2} \right\}\nonumber\\
&\times&\prod_i \frac{1}{\sqrt{2\pi}\sigma_{f, i}}\exp\left\{ - \frac{[f_i-f_i^m]^2}{2\sigma_{f, i}^2} \right\},
\label{eqn_psa}
\end{eqnarray}
where the subscript $i$ represents $i$-th velocity bin, the superscript $m$ represents the modeled quantity,  $\Delta\phi_i$ and $\sigma_{\phi, i}$ are the differential phase and uncertainty, and $f_i$ and $\sigma_{f, i}$ are the line flux density and uncertainty, respectively.
The RM data include the H$\beta$ and continuum light curves. The corresponding likelihood function is given by
\begin{equation}
\mathcal{L}_{\rm RM}=\prod_i \frac{1}{\sqrt{2\pi}\sigma_{F, i}}\exp\left\{ - \frac{[F_i-F_i^m]^2}{2\sigma_{F, i}^2} \right\},
\label{eqn_prm}
\end{equation}
where the subscript $i$ represents $i$-th epoch, the superscript $m$ represents the modeled quantity, and $F_i$ and $\sigma_{F, i}$ are the H$\beta$ line flux and uncertainty, respectively. Here, the modeled line flux $F_i^m$ depends on the continuum light curve modeling.

The joint likelihood function of SARM analysis is the product of Equations~(\ref{eqn_psa}) and (\ref{eqn_prm}), namely, $\mathcal{L}_{\rm SARM} = \mathcal{L}_{\rm SA}\times \mathcal{L}_{\rm RM}$. The posterior probability distribution of model parameters $\bm{\Theta}$ is then $P(\bm{\Theta}|D_{\rm SARM}) = P(\bm{\Theta})\times \mathcal{L}_{\rm SARM}$, where $P(\bm{\Theta})$ is the parameter prior. In Appendix~\ref{app_parameters}, we list the full model parameters and their priors.
We employ the diffusive nested sampling (DNS) algorithm (\citealt{Brewer2011}) to explore the parameter space and generate Markov chains, which has been shown to be effective at exploring multimodal distributions and strong correlations between parameters. We have incorporated the above BLR dynamical modeling into our previously developed package \textsc{BRAINS}\footnote{The latest code is publicly available at \url{https://github.com/LiyrAstroph/BRAINS}, while the version used in this work is available in \cite{brains}.} (\citealt{Li2013, Li2018}), which utilizes the DNS code \textsc{CDNest}\footnote{The latest code is publicly available in \url{https://github.com/LiyrAstroph/CDNest}, while the version used in this work is available in \cite{cdnest}.}.
Both \textsc{BRAINS} and \textsc{CDNest} are written in C Language with the standardized message passing interface and are thereby portable to a wide range of supercomputer clusters. From the generated posterior samples of parameters, we assign the best estimates by the median values and the uncertainties by the 68.3\% confidence interval.

\section{Results}\label{sec_results}
\subsection{Overview}
In Figures~\ref{fig_Mrk509}-\ref{fig_NGC3783}, we show the SARM analysis results of the four AGNs, including fits to the SA and RM data, and posterior distributions of the BLR size, black hole mass, and angular-diameter distance. Table~\ref{table_key_para} lists the best estimates and uncertainties of these three key parameters, together with the BLR angular size. A full list of the inferred parameter values and uncertainties are summarized in Appendix~\ref{app_parameters}.
For the sake of clarity, we only plot the differential phases averaged over the three baselines with the strongest BLR phase signals. The continuum phases have also been subtracted. The best fits to the individual differential phases of these three baselines are presented in Appendix~\ref{app_sa}.

When running the MCMC, we find that for PDS 456, the continuum phase offset parameter is strongly degenerate with the angular-diameter distance. The Bayesian modeling always tends to fit the whole differential phases with the continuum phase, leaving quite small phase residuals for the BLR emissions. This in turn results in an unrealistically large angular-diameter distance. We therefore switch off the continuum phase offset parameters for PDS 456. As illustrated below, the distance obtained without including the continuum phase offset turns out to be more reasonable. In addition, we also note that the H$\beta$ light curve, shown in the left bottom panel of Figure~\ref{fig_PDS456}, displays some fine variation patterns that do not have counterparts in the continuum light curve. It seems that these non-echoing H$\beta$ variations are not solely due to noise but their origin is unclear. The analysis reproduces only the averaged variation trend of the H$\beta$ light curve. As a result, there appear a number of data points deviating from the best reconstruction.

In the averaged differential phase curves (with the continuum phase subtracted) plotted in the second panel of the top left row of Figures~\ref{fig_Mrk509}-\ref{fig_NGC3783}, 3C~273 and NGC~3783 display an approximately S-shaped signal typical for Keplerian rotation, which is characterized by a crest and trough respectively on either side of the center wavelength of the emission line. In contrast, PDS~456 and Mrk~509 show an asymmetric signal dominated by a trough in the differential phase curve, indicative of non-Keplerian motion. In the present dynamical modeling, the latter can be explained by the presence of a fraction of BLR clouds with inflowing/outflowing motion plus anisotropic emissions. For PDS 456, there appears an extra red-wing component around 2.24~$\mu$m, which is not reproduced by our current model. We emphasize that the plotted differential phase curve indeed represents the averaged data from the three baselines with the strongest BLR phase signals (see Appendix C). Notably, this component does not appear across all baselines, possibly implying that it might be caused by some unknown measurement noise or a real BLR component at a special location and velocity that only produces signals along certain baselines. With the available quantity and quality of data we cannot discriminate between these two possibilities. Future higher-quality data will facilitate either a more advanced BLR modeling or a confirmation that we have here a transient observing feature if the extra red-wing component is real.

\subsection{The Inferred Radial-dependent Responsivity}
Figure~\ref{fig_eta} illustrates the obtained responsivity as a function of radius, along with the 1$\sigma$ confidence intervals. We find that 3C 273 and PDS 456 show a strong radial-dependent responsivity. In contrast, the dependence for Mrk 509 and NGC 3783 is relatively weak. For a strongly radially increasing responsivity, we expect a significant difference in the emission-line widths from the mean and rms spectra in the RM data (see \citealt{LW2024}). This is because the BLR clouds are more responsive at greater radii with a lower velocity. These clouds therefore add larger weights to variations of the line flux density at low-velocity bins, resulting in a narrower line width in rms spectra compared to the mean spectra.
In earlier studies in literature, two types of line width are used, the full width at half maximum (FWHM) and the line dispersion ($\sigma_{\rm line}$). In Appendix~\ref{app_spec}, we show the mean and rms spectra from previous RM campaigns for the four AGNs. For 3C 273, the H$\beta$ FWHM in the rms spectrum ($\sim1900~\rm km~s^{-1}$) is narrower than in the mean spectrum ($\sim3300~\rm km~s^{-1}$)  by a factor of $\sim$0.4 (\citealt{Zhang2019}). The same is the case for the H$\beta$ $\sigma_{\rm line}$. Similar to 3C~273, PDS~456 also has significantly narrower H$\beta$ line widths in the rms spectrum (\citealt{Li2024}). In contrast, the H$\beta$ line widths of Mrk 509 (\citealt{Li2024}) and NGC 3783 (\citealt{Bentz2021}) are similar for the mean and rms spectra. Our obtained responsivities in Figure~\ref{fig_eta} are remarkably coincident with the difference between the mean and rms H$\beta$ line widths, implying a validation of our dynamical modeling. The responsivity can also be further tested by future photoionization calculations.

The responsivity directly affects the variation amplitude of the emission line. From Equation~(\ref{eqn_fl}), we can deduce that the responsivity tends to scale up ($\eta>1$) or down ($\eta<1$) the variation amplitude of the emission line relative to that of the continuum. As mentioned in Section~\ref{sec_rmdata}, in real observational data the continuum flux also contains the extra contribution from the host galaxy's starlight or other components, causing contamination to the constant continuum component.  This in turn requires a larger responsivity to reproduce the fractional variability amplitude of the emission line. As a result, the inferred responsivity will be biased to a larger value in the presence of other contamination. In our analysis, we have corrected for the contamination by the host galaxy for NGC~3783 and the jet emission for 3C~273. The contamination of the host galaxy for the other two objects Mrk 509 and PDS 456 may be treated as negligible. We also run SARM analysis with the original uncorrected RM data and confirm that these corrections do not affect the obtained geometric distances. 

In our analysis, we set an upper limit of the responsivity of $\eta_{\rm max}=1.5$ (see Equation~\ref{eqn_eta}), based on previous photoionization calculations (e.g., \citealt{Korista2004, Goad2014, Goad2015, Zhang2021, LW2024}). In Appendix~\ref{app_etamax}, we rerun analysis with two different upper limits of $\eta_{\rm max}=1.2$ and $2$ and illustrate the influences on the BLR size, SMBH mass, and angular-diameter distance. We find that the results for Mrk~509, PDS~456, and NGC~3783 are quite insensitive to the choice of $\eta_{\rm max}$. For 3C~273, the best estimates of the BLR size and angular-diameter distance are relatively smaller for $\eta_{\rm max}=2$, but still consistent with those of $\eta_{\rm max}=1.2$ and $1.5$ within 1$\sigma$ uncertainties.

%=============================================================================================
\begin{figure*}[th!]
\centering
\includegraphics[width=0.9\textwidth]{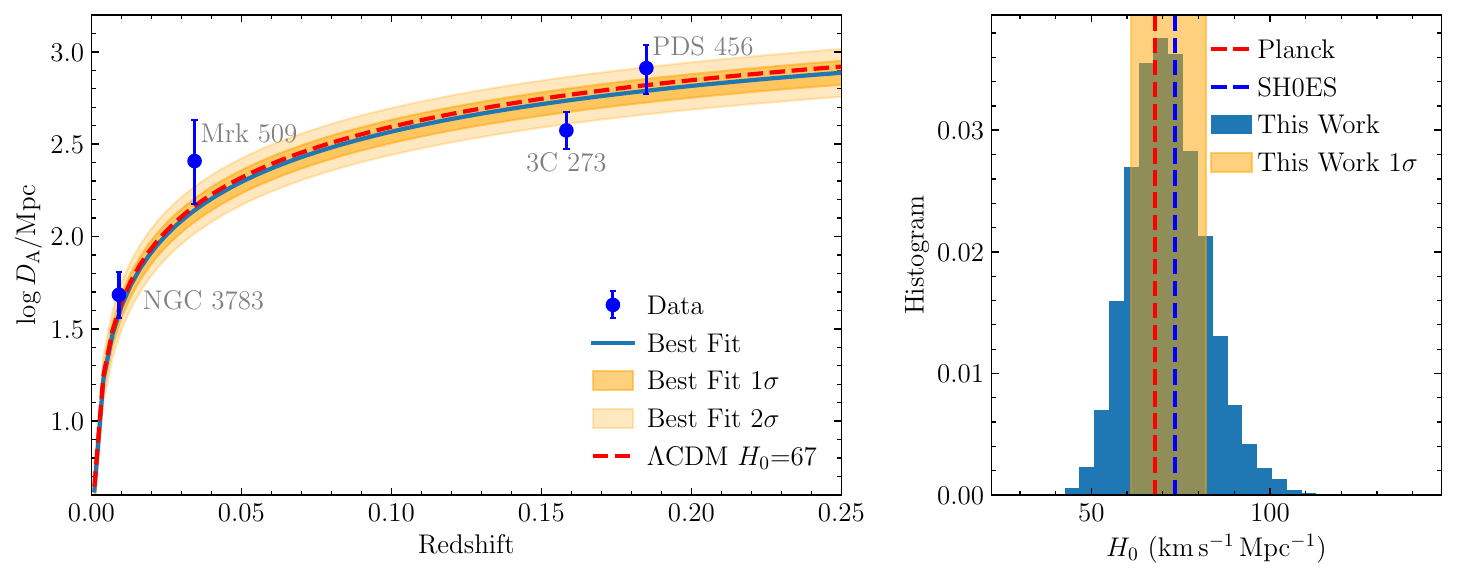}
\caption{(Left) The obtained geometric distances with redshifts of the four AGNs. The solid blue line represents the best fit to the distances using a flat $\Lambda$CDM cosmology and the dark and light orange shaded bands represent the 1$\sigma$ and 2$\sigma$ uncertainties, respectively. The red dashed line corresponds to $H_0=67~\rm km~s^{-1}~Mpc^{-1}$. The redshift of NGC 3783 has been corrected for the peculiar velocity (see Section~\ref{sec_h0}). (Right) The posterior distribution of the Hubble constant. The vertical orange shaded band represents the 1$\sigma$ uncertainties. The vertical red and blue lines represent the Hubble constant measured from the Planck satellite (\citealt{Planck2020}) and the SH0ES project (\citealt{Reid2019}). In deriving the Hubble constant, the other parameters of the flat $\Lambda$CDM cosmology are fixed to $\Omega_{\rm M}=0.3$ and $\Omega_{\Lambda}=0.7$.}
\label{fig_DA}
\end{figure*}
%=============================================================================================

\begin{deluxetable*}{lcccc}
\tablecaption{Inferred Values of the Key Parameters and Their Uncertainties.\label{table_key_para}}
\tablehead{
\colhead{Parameter}  & \colhead{Mrk 509} & \colhead{PDS 456}  & \colhead{3C 273}  & \colhead{NGC 3783}
}
\startdata
$\log(D_{\rm A}/\rm Mpc)$ & $2.41_{-0.23}^{+0.22}$ & $2.91_{-0.13}^{+0.13}$ & $2.58_{-0.11}^{+0.09}$ & $1.68_{-0.13}^{+0.12}$ \\
$\log(M_\bullet/M_\odot)$ & $7.86_{-0.23}^{+0.24}$ & $7.76_{-0.14}^{+0.23}$ & $8.55_{-0.15}^{+0.21}$ & $7.17_{-0.05}^{+0.07}$ \\
$R_{\rm BLR}/\rm ld$ & $106.46_{-42.84}^{+94.09}$ & $144.13_{-39.22}^{+79.89}$ & $84.70_{-8.25}^{+10.73}$ & $24.90_{-3.16}^{+4.12}$ \\
$\Theta_{\rm BLR}/\rm \mu as$ & $73.89_{-28.42}^{+53.11}$ & $31.72_{-8.03}^{+12.96}$ & $39.60_{-7.98}^{+9.05}$ & $89.19_{-20.64}^{+28.23}$ \\
\enddata
\tablecomments{The BLR angular size is defined by $\Theta_{\rm BLR}=R_{\rm BLR}/D_{\rm A}$.}
\end{deluxetable*}

\subsection{Geometric Distances}
The geometric distances of Mrk~509 and PDS 456 are measured here for the first time, while the analysis of the other two sources, 3C 273 and NGC 3783, had been conducted by \citetalias{Wang2020} and \citetalias{Gravity2021b}, respectively, based on linear BLR responses. Below we first present our distance measurements of Mrk 509 and PDS 456, and then our latest measurements of  3C~273 and NGC~3783, together with a detailed comparison with the previous results. The left panel of Figure~\ref{fig_DA} shows the distribution of the four quasars in the plane of redshift versus angular-diameter distance.

We obtain a distance of $\log(D_{\rm A}/\rm Mpc)=2.41_{-0.23}^{+0.22}$ for Mrk 509 and $2.91_{-0.21}^{+0.14}$ for PDS 456. For comparison, if using a flat $\Lambda$CDM cosmology with $H_0=67~\rm km~s^{-1}~Mpc^{-1}$, $\Omega_{\rm M}=0.3$, and $\Omega_\Lambda=0.7$, the corresponding angular-diameter distances are $\log(D_{\rm A}/\rm Mpc)=2.17$ and 2.83, respectively. Those values are compatible with our inferences within 1$\sigma$ uncertainties.

Our inferred angular-diameter distance of 3C 273 is $\log(D_{\rm A}/\rm Mpc)=2.58_{-0.11}^{+0.09}$ and of NGC~3783 is $1.70_{-0.13}^{+0.13}$. The corresponding distances estimated from the cosmology are 2.77 and 1.63, respectively, again compatible with our inferences within 2$\sigma$ uncertainties. NGC 3783 is quite nearby and therefore subject to non-negligible peculiar motion (see \citetalias{Gravity2021b} for a detailed investigation). If using a peculiar velocity $v_{\rm pec}\sim-158~\rm km~s^{-1}$ (see the next section for details), the corrected redshift will be smaller by about 5\%. The anticipated $D_{\rm A}$ for NGC 3783 from cosmology will be reduced by the same factor, to $\log(D_{\rm A}/{\rm Mpc})=1.61$. In any case, our obtained geometric distances are generally consistent with the cosmology-based estimates.

Using SARM analysis, \citetalias{Wang2020} derived $\log(D_{\rm A}/\rm Mpc)=2.74_{-0.07}^{+0.08}$ for 3C 273 and \citetalias{Gravity2021b} derived  $1.60_{-0.15}^{+0.13}$ for NGC~3783. Considering the associated uncertainties of the inferred distances, these values are consistent with our results. However, we stress that the BLR dynamical modeling in both \citetalias{Wang2020} and \citetalias{Gravity2021b} adopted a globally linear, uniform responsivity, which always results in the same emissivity- and responsivity-weighted BLR sizes (see also \citealt{LW2024}). Our improved SARM analysis naturally accounts for the differences in these two BLR sizes and largely alleviates the increased systematic biases in distance measurements. As mentioned in the preceding section, we find that 3C~273 shows a strong radial dependence of the responsivity, while this dependence is weak in NGC~3783 (see Figure~\ref{fig_eta}).  These results remarkably coincide with the line widths in the mean and rms spectra of the two objects (see Appendix~\ref{app_spec}).

For 3C~273, the smaller center value of $D_{\rm A}$ by $\sim$0.16 dex we obtained compared to the one given by \citetalias{Wang2020} can be understood as follows. Due to the use of linear responses in \citetalias{Wang2020}, the emissivity-weighted BLR size of 3C~273 tends to be overestimated and the resulting distance is therefore forced to be larger given the same differential phase data. For NGC~3783, \citetalias{Gravity2021b} demonstrated that an inclusion of non-linear responses (still globally uniform) results in a distance about 10\% larger. Note that we included the photometric light curve from the ASAS-SN archival data, which significantly extends the temporal baseline of the continuum light curve and therefore enables better continuum modeling. For example, the spectral monitoring of \cite{Bentz2021} started on JD2,458,891, while ASAS-SN database includes $g$-band photometry since JD2,458,812, which is 79 days earlier (see Figure~\ref{fig_NGC3783}). The above two factors might explain the reason why our inferred distance is larger by 0.1 dex than that of \citetalias{Gravity2021b}, although both are still consistent within uncertainties.

The statistical uncertainties in $D_{\rm A}$ of the four objects range from 0.09 to 0.24 dex, stemming from the uncertainty of the linear size (constrained by the RM data) and angular size (constrained by the SA data) of the BLR. The inferred values in Table~\ref{table_key_para} imply that
the uncertainties of $D_{\rm A}$ mainly come from the SA data for 3C 273 and NGC 3783, whereas for Mrk 509 and PDS 456, the RM and SA data make comparable contributions to the uncertainties.

%=============================================================================================
\begin{figure*}[ht!]
\centering
\includegraphics[width=0.95\textwidth]{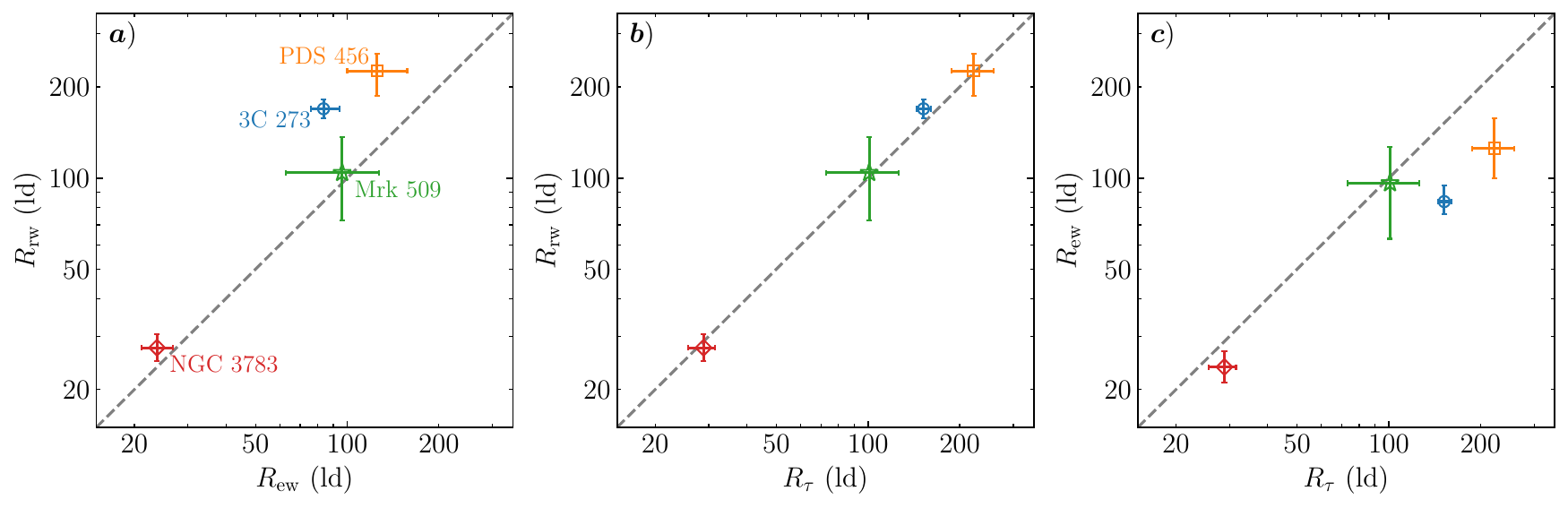}
\caption{A comparison between three measures of BLR sizes of the four quasars, the responsivity-weighted size ($R_{\rm rw}$), the emissivity-weighted size ($R_{\rm ew}$), and the size  $R_{\tau}$ from the centroid time delay of the responsivity-weighted transfer function. The panels are a) $R_{\rm rw}$ vs. $R_{\rm ew}$, b) $R_{\rm rw}$ vs. $R_{\tau}$, and c) $R_{\rm ew}$ vs. $R_{\tau}$. The grey dashed line represents the line of equality. See Appendix~\ref{app_size} for the definitions of these BLR sizes. We stress that $R_{\rm ew}$, $R_{\rm rw}$, and $R_{\tau}$ represent different measures of BLR sizes. The comparisons presented here by no means indicate that SA underestimates the BLR sizes or RM overestimates the BLR sizes.}
\label{fig_size}
\end{figure*}

\begin{figure}[ht!]
\centering
\includegraphics[width=0.33\textwidth]{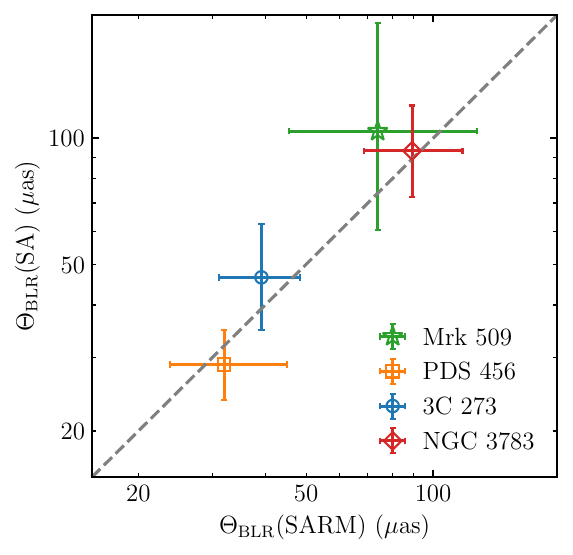}
\caption{A comparison of angular BLR sizes $\Theta_{\rm BLR}=R_{\rm BLR}/D_{\rm A}$ obtained from SARM analysis
and sole SA analysis. }
\label{fig_size_sa}
\end{figure}

%=============================================================================================

\subsection{An Estimate of the Hubble Constant}\label{sec_h0}
The measured angular-diameter distances allow us to estimate the Hubble constant. For a flat $\Lambda$CDM cosmology, the relation between redshift and angular-diameter distance
is
\begin{equation}
D_{\rm A}(z) = \frac{c}{(1+z)H_0}\int\frac{dz}{\sqrt{\Omega_\Lambda + \Omega_{\rm M}(1+z)^3}},
\label{eqn_DA}
\end{equation}
where $c$ is the speed of light, $H_0$ is the Hubble constant, $\Omega_{\rm M}$ is the dimensionless density of mass, and the dimensionless density of dark energy $\Omega_\Lambda=1-\Omega_{\rm M}$.
Because of the rather limited redshift span of the four sources, the parameter $\Omega_{\rm M}$ can not be constrained properly and we simply adopt $\Omega_{\rm M}=0.3$. Note that this value has little effect on the determination of the Hubble constant.
For NGC 3783, according to the comprehensive investigation by \citetalias{Gravity2021b}, the peculiar velocity was estimated to be $v_{\rm pec}=-158~\rm km~s^{-1}$ from the averaged peculiar velocity of 11 galaxies  within 8$h^{-1}$ Mpc centered on the position of NGC 3783 obtained from the 6dF galaxy redshift survey peculiar velocity map (\citealt{Springob2014}).
We correct the redshift of NGC 3783 by taking into account the peculiar velocity as (\citealt{Davis2014})
\begin{equation}
z_{\rm cor} = z + \frac{v_{\rm pec}}{c} (1 + z),
\end{equation}
where $z$ is the normal recession redshift. This leads to $z_{\rm cor}=0.009198$.

By fitting the inferred angular-diameter distances of four AGNs with Equation~(\ref{eqn_DA}), we obtain a Hubble constant of $H_0=69_{-10}^{+12}~\rm km~s^{-1}~Mpc^{-1}$.
In the right panel of Figure~\ref{fig_DA}, we plot the posterior distribution of $H_0$, along with the measurements from the Planck satellite (\citealt{Planck2020}) and SH0ES project (\citealt{Reid2019}). Our estimated Hubble constant bears a large uncertainty ($\sim$15\%) that is insufficiently precise to address the current Hubble tension problem (e.g., \citealt{Riess2020}). However, the forthcoming upgraded instrument GRAVITY+ will significantly improve the sensitivity magnitude down to $K\sim13$,
allowing observations of a few hundred AGNs with higher-quality differential phase measurements (\citealt{Gravity+2022}), which will enable the SARM approach to achieve a competitive percent-level precision (\citealt{Songsheng2021}).

%=============================================================================================
\begin{figure*}
\centering
\includegraphics[width=0.4\textwidth]{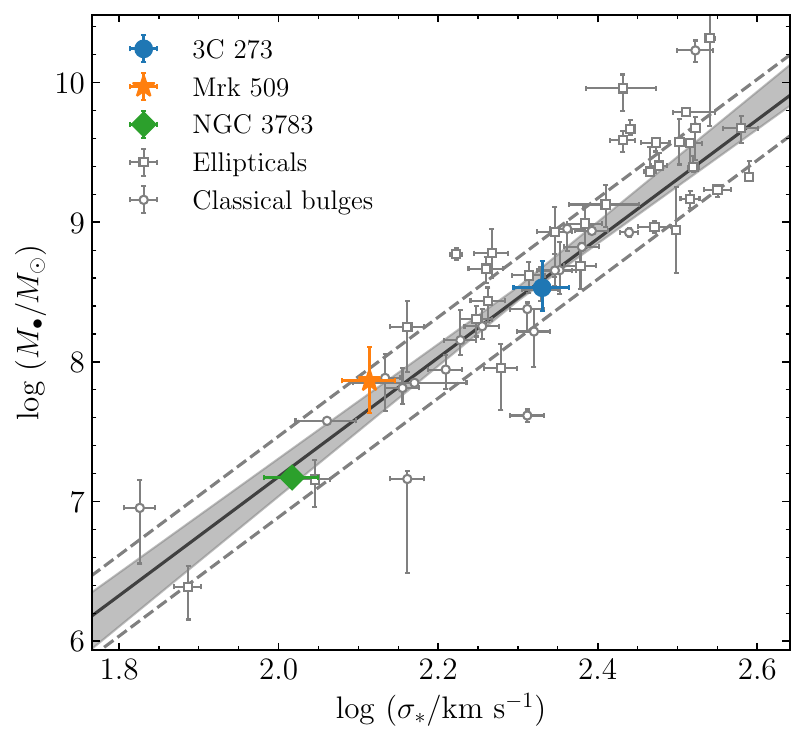}
\includegraphics[width=0.4\textwidth]{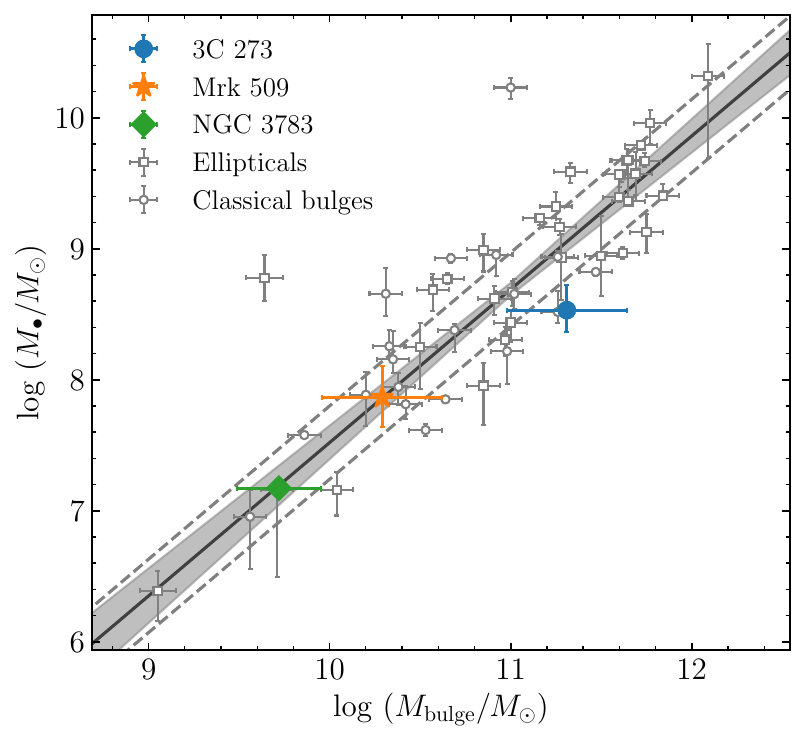}
\caption{The location of Mrk 509, 3C 273, and NGC 3783 on the plots between SMBH masses and  (left) stellar velocity dispersions and (right)  masses of the bulges.
The gray open points are the data compiled by \cite{Kormendy2013}. Solid lines with shaded areas represent the linear regression fits and gray dashed lines represent the intrinsic scatters of the relations reported by \cite{Kormendy2013}. The stellar velocity dispersions and masses of the bulges for the three quasars are from \cite{Winkel2024}. There are not yet such measurements for PDS~456.}
\label{fig_relation}
\end{figure*}
%=============================================================================================

\subsection{BLR Sizes and SMBH Masses}
As mentioned above, the radial-dependent responsivity results in different responsivity- and emissivity-weighted BLR sizes (denoted as $R_{\rm rw}$ and $R_{\rm ew}$, respectively). Moreover, owing to the inclusion of anisotropic effects of BLR emissions (see Section~\ref{sec_model}), the centroid time delay $R_\tau$ of the responsivity-weighted transfer function $\Psi_r(\tau)$, which is a quantity directly related to RM analysis, might not be exactly the same as $R_{\rm rw}$.
In Appendix~\ref{app_size}, we illustrate how to calculate these BLR sizes given a BLR dynamical model.
Figure~\ref{fig_size} shows comparisons among the obtained BLR sizes for the four sources. As expected, $R_{\rm rw}$ is larger than $R_{\rm ew}$ by $\sim60\%$ for PDS 456 and 3C 273, both showing a significant radially increasing responsivity (see Figure~\ref{fig_eta}). There appears a good consistency between $R_{\rm rw}$ and $R_\tau$ because the transfer function $\Psi_r(\tau)$ is closely related to the responsivity as illustrated in Equation~(\ref{eqn_psir}). As a consequence, the size $R_{\rm ew}$ of both PDS~456 and 3C~273 is smaller than $R_\tau$ by a factor of $\sim$60\%. These results imply that when comparing the BLR sizes obtained from interferometry (corresponding to $R_{\rm ew}$) against those from RM (corresponding to $R_{\tau}$), one needs to carefully take into account the differences in the BLR dimensions represented by $R_{\rm ew}$ and $R_{\rm \tau}$. In other words, Figure~\ref{fig_size} by no means indicates that GRAVITY interferometry underestimates the BLR sizes or RM overestimates the BLR sizes. Indeed, these two approaches probe different BLR measures. To further illustrate the above point, we apply the same BLR dynamical model to only GRAVITY data and make a comparison of the obtained angular sizes ($\Theta_{\rm BLR}=R_{\rm BLR}/D_{\rm A}$) with the sizes from our SARM analysis in Figure~\ref{fig_size_sa}. Here, $R_{\rm ew}$ is used to calculate the angular sizes in SARM analysis. We can find a good agreement between the two approaches. This in turn means that when using $R_{\rm rw}$ to calculate the SARM angular size, there will be a significant difference from the SA angular size for PDS~456 and 3C~273.

For 3C 273, \cite{Gravity2018} measured a BLR angular size of $46\pm10$ $\mu$as, consistent with our result of $39.60_{-7.98}^{+9.05}$ $\mu$as (see Table~\ref{table_key_para}). However, we note that the linear BLR size of $145\pm35$ ld reported by \cite{Gravity2018} is larger than our inference of $84.70_{-8.25}^{+10.73}$ ld. This is because they adopted  a distance of 550 Mpc, larger than our measured distance of $380_{-85}^{+86}$ Mpc. For the other three sources, we can also find an agreement with previous SA analysis within uncertainties (\citealt{Gravity2021a, Gravity2024}).

We obtain a black hole mass $\log(M_\bullet/M_\odot)=7.86_{-0.23}^{+0.24}$ for Mrk 509, $7.76_{-0.14}^{+0.23}$ for PDS 456,  $8.55_{-0.15}^{+0.21}$ for 3C 273, and $7.17_{-0.05}^{+0.07}$ for NGC 3783. Recently, using high-spectral-resolution integral-field spectroscopic observations, \cite{Winkel2024} measured the stellar velocity dispersions and dynamical masses of the host bulges for a sample of nearby AGNs, including Mrk~509, 3C~273, and NGC~3783.
In Figure~\ref{fig_relation}, combined with our obtained SMBH masses, we superimpose these three objects on the scaling correlations between SMBHs and bulges compiled by \cite{Kormendy2013}.  We find that the three objects match these correlations within the associated uncertainties.

By comparing with SMBH mass determinations using only RM, our inferred masses, except for PDS 456, are generally consistent with previous RM results within uncertainties (\citealt{Zhang2019, Bentz2021, Li2024}).
When using a unity virial factor and the H$\beta$ FWHM from the mean spectrum, the RM campaign of \cite{Li2024} reported a mass of $\log(M_\bullet/M_\odot)=8.59_{-0.11}^{+0.07}$ for PDS 456. Note that such a RM-based mass estimate does not include the scatter of the virial factor ($\sim$0.3 dex; e.g., \citealt{Ho2014}), hence underestimating the uncertainties. The optical spectrum of PDS~456 shows strong iron emissions and the [\ion{O}{3}]$\lambda\lambda$4959, 5007 lines are almost invisible (see Appendix~\ref{app_spec}). These features are commonly seen in narrow-line Seyfert I galaxies and are believed to be indicative of super-Eddington accretion (e.g., \citealt{Boroson1992, Collin2004, Du2014}). Thus a lower SMBH mass might be more plausible. Future comprehensive observations with improved data quality in both SA and RM are highly worthwhile to verify the present black hole mass estimates.

Compared to the SMBH mass inferred from the GRAVITY data alone (\citealt{Gravity2018, Gravity2021a, Gravity2024}), we again find overall consistency within 1$\sigma$ uncertainties. In particular for PDS~456, the SA analysis by \cite{Gravity2024} obtained an SMBH mass of $\log(M_\bullet/M_\odot)=8.23_{-0.49}^{+0.01}$. Despite this value differing from our result by 0.47 dex, the lower error of $0.49$ dex implies that the two results are statistically in agreement.

%=============================================================================================
\begin{figure*}[hpt!]
\centering
\includegraphics[width=0.95\textwidth]{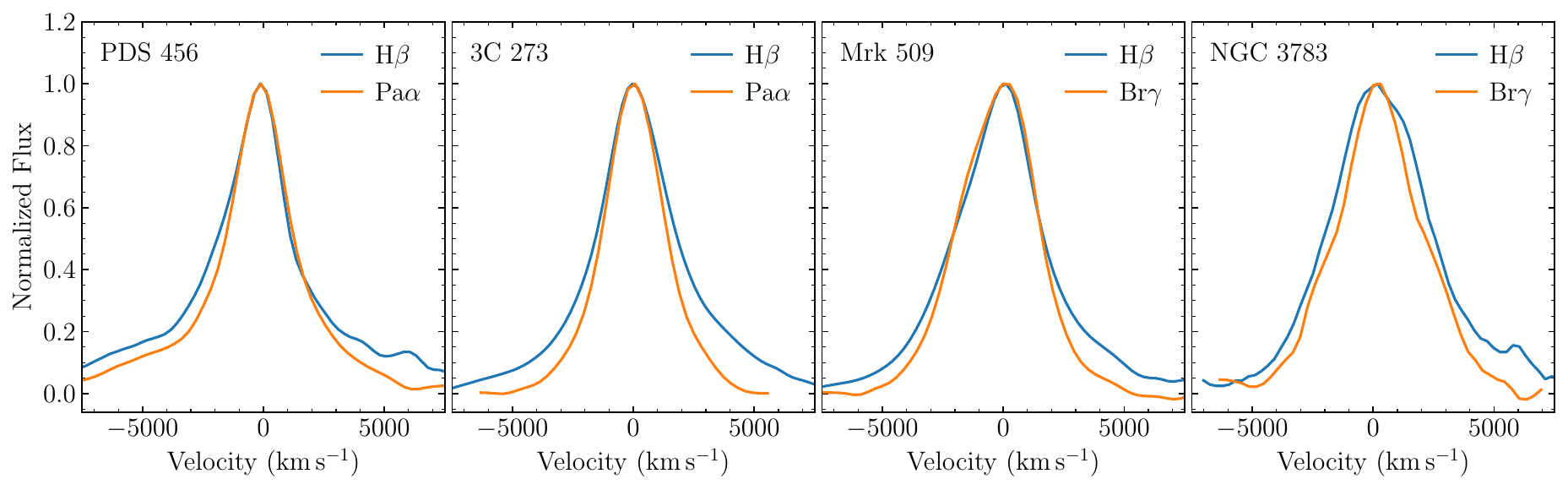}
\caption{A profile comparison between the Pa$\alpha$/Br$\gamma$ line observed in SA and the H$\beta$ line observed in RM. The Pa$\alpha$/Br$\gamma$ lines are manually convolved with a Gaussian to match the spectral resolution of the H$\beta$ line (\citealt{Zhang2019, Bentz2021, Li2024}). The Gaussian standard deviation is adopted to $\sim$400 km~s$^{-1}$ for PDS 456, 3C 273, and Mrk 509, and to $\sim$230 km~s$^{-1}$ for NGC 3783. The original spectral resolution of the Pa$\alpha$/Br$\gamma$ lines is $\sim$235 km~s$^{-1}$. }
\label{fig_profile}
\end{figure*}
%=============================================================================================
%=============================================================================================
\begin{figure*}[hpt!]
\centering
\includegraphics[width=0.95\textwidth]{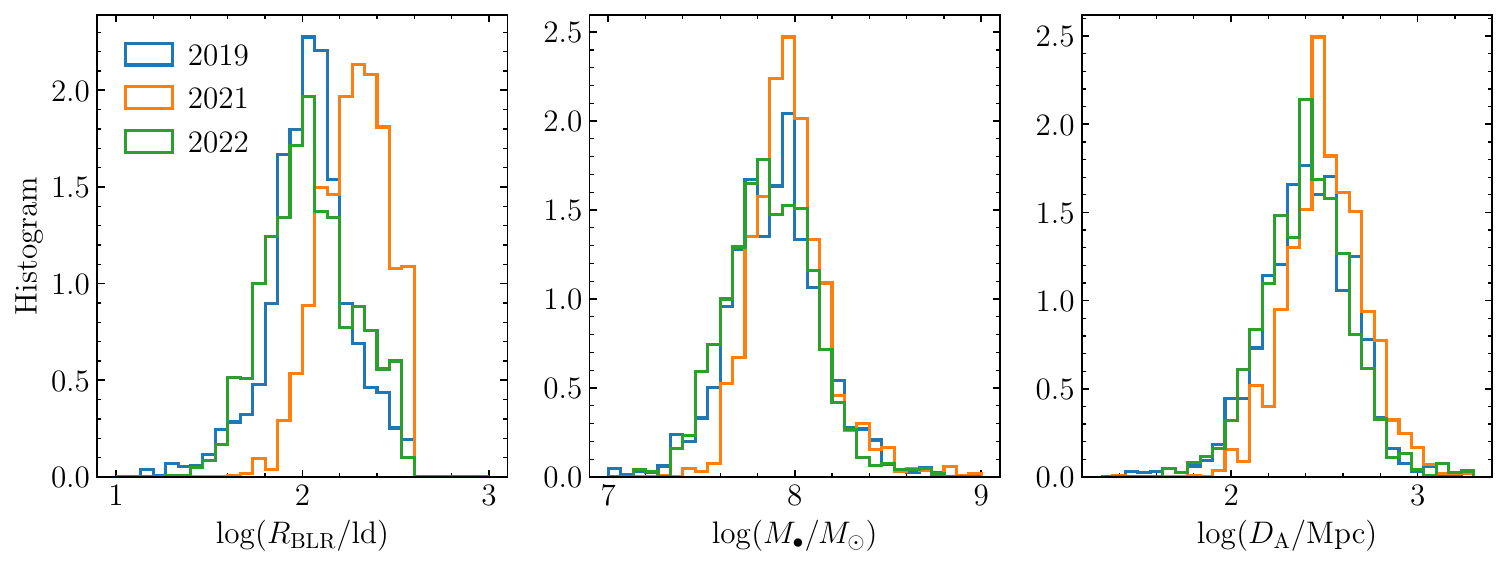}
\caption{A comparison of (left) the obtained BLR size, (middle) the SMBH mass, and (right) the angular-diameter distance of Mrk 509 using the RM data in 2019, 2021, and 2022.}
\label{fig_comp_mrk509}
\end{figure*}
%=============================================================================================

\section{Discussion}\label{sec_dis}
\subsection{Differences between the H$\beta$ and Pa$\alpha$/Br$\gamma$ Lines}
An implicit assumption in the present SARM analysis is that the optical H$\beta$ line probed by RM and the near-infrared Pa$\alpha$/Br$\gamma$
line probed by SA stem from exactly the same region. In practice, the observed profiles of these lines are slightly different. Figure~\ref{fig_profile} compares the Pa$\alpha$/Br$\gamma$ line with the H$\beta$ line subtracted from the mean spectra of the RM campaign, all normalized by the peak flux density. The Pa$\alpha$/Br$\gamma$ lines are also manually convolved with a Gaussian to match the spectral resolution of the H$\beta$ lines. As observed, the Pa$\alpha$/Br$\gamma$ line is overall narrower than H$\beta$. After subtracting line width broadening arising from the spectral resolution, we estimate the line widths as follows: FWHM(H$\beta$, Pa$\alpha$)=(2840, 2620) km~s$^{-1}$ for PDS~456, FWHM(H$\beta$, Pa$\alpha$)=(3220, 2580) km~s$^{-1}$ for 3C 273, FWHM(H$\beta$, Br$\gamma$)=(3650, 3560) km~s$^{-1}$ for Mrk 509, and FWHM(H$\beta$, Br$\gamma$)=(4340, 3680) km~s$^{-1}$ for NGC 3783 (see also \citealt{Gravity2021a}). The resulting line width ratio between the H$\beta$ and Pa$\alpha$/Br$\gamma$ lines is 1.09, 1.25, 1.02, and 1.18, respectively. Using the virial relation $R\propto V^{-2}$, the corresponding differences in the BLR sizes of the two emission lines become 18\%, 56\%, 5\%, and 39\%, respectively.

The fact that the H$\beta$ line is relatively broader means that current SARM analysis underestimates the geometric distance. If simply scaling results with the above estimated differences in the BLR sizes, the geometric distance will increase by roughly 0.07, 0.19, 0.02, and 0.14 dex for PDS~456, 3C~273, Mrk~509, and NGC~3783, respectively. However, we note that a precise evaluation of the influences on the distance measurements requires comprehensive modeling of the H$\beta$ and Pa$\alpha$/Br$\gamma$ lines simultaneously, which is beyond the scope of the present work. Nevertheless, our estimate of the Hubble constant remains within a reasonable range, indicating that the
discrepancy arising from different emission lines measured by SA and RM is moderate. In the future, this issue can be surmounted by near-infrared RM campaigns observing the same broad emission line as SA or a joint analysis of SA and velocity-resolved RM. In the latter approach, one no longer needs to assume the same region for the H$\beta$ and Pa$\alpha$/Br$\gamma$ lines but only let the two lines share the same central SMBH (\citealt{Li2022}).

\subsection{Results of Mrk 509 from Different Years of RM Data}
Mrk 509 was monitored for four years between 2019 and 2022 and the H$\beta$ time delay can be detected in three years, except for 2020 due to the absence of clear variation patterns in the RM data (\citealt{Li2024}). We chose the 2022 RM data for SARM analysis given that this year marked the largest variability. In Figure~\ref{fig_comp_mrk509}, we illustrate a comparison of the obtained BLR size, SMBH mass, and angular-diameter distance between the years 2019, 2021, and 2022. As observed, the best estimates are consistent with each other within uncertainties, although the BLR size in 2021 shows a slightly larger median value. This is as expected since the H$\beta$ time delays in these three years are in overall agreement.

\section{Conclusions}\label{sec_conclusion}
We combine SA and RM data to perform joint SARM analysis to measure the geometric distances of four AGNs: Mrk 509, PDS 456, 3C 273, and NGC 3783. SARM results are reported for the first time for the former two sources and are revisited for the latter two. Inspired by photoionization calculations and the long-standing observation from RM campaigns that H$\beta$ emission-line widths in rms spectra are generally narrower than those in mean spectra,  we improve BLR dynamical modeling by including a radially dependent responsivity. This enables a self-consistent treatment of the emissivity weighting of the BLR in SA and responsivity weighting in RM and thereby eliminates any potential biases that may arise from imposing a uniform, linear BLR response. Such a non-linear, radially dependent responsivity has been neglected in all previous dynamical modeling approaches. We obtain an angular-diameter distance of $\log(D_{\rm A}/{\rm Mpc})=2.40_{-0.30}^{+0.19}$ for Mrk 509, $2.91_{-0.13}^{+0.13}$ for PDS 456, $2.58_{-0.09}^{+0.11}$ for 3C 273, and $1.68_{-0.13}^{+0.12}$ for NGC 3783, from which we derive a Hubble constant of $H_0=69_{-10}^{+12}\,\rm km\,s^{-1}\,Mpc^{-1}$.
The SARM analysis also allow us to measure a black hole mass $\log(M_\bullet/M_\odot)=7.86_{-0.23}^{+0.24}$ for Mrk 509, $7.76_{-0.14}^{+0.23}$ for PDS~456, $8.55_{-0.15}^{+0.21}$ for 3C 273, and $7.17_{-0.05}^{+0.07}$ for NGC~3783. These values are in good agreement with known  relations between black hole masses and stellar velocity dispersions or dynamical masses of the host bulges (though for PDS 456 without relevant measurements of the host galaxy).

The currently observed AGNs by GRAVITY remain rare due to the strict magnitude limit ($K\lesssim10$; \citealt{Gravity2017}). However, the forthcoming upgraded instrument GRAVITY+ will enable phase reference on a target with $K<13$ using the same observation mode (``on-axis''\footnote{Alternatively, with the ``off-axis'' mode, fainter AGNs ($K>13$) are observable provided there is a bright star ($K<13$) within a 30{\arcsec} separation serving as the phase reference. However, due to reduced photon fluxes, these AGNs require extended exposure times to achieve an adequate signal-to-noise ratio for BLR differential phases observations.}) as the targets reported here. This will permit interferometric observations of a much larger number of AGNs (up to a few hundred; \citealt{Gravity+2022}) and thereby can significantly expand the AGN sample for SARM analysis. The improved sensitivity of GRAVITY+ can also facilitate higher-quality differential phase measurements. Given the feasibility of RM monitoring, those $z<0.3$ AGNs observable with the on-axis mode are vital for geometric distance determinations.  In particular, the Pa$\alpha$ line, redshifted into the $K$ band for $0.1 \lesssim z \lesssim 0.3$, is generally about 10 times brighter than the Br$\gamma$ line for $z\lesssim 0.1$. Therefore, the Pa$\alpha$ AGN sample is preferred for future SARM observations with GRAVITY+. In addition, a new visitor interferometry instrument BIFROST at the $J$ and $H$ bands on VLTI will come into operation within the next few years (\citealt{Kraus2022}). It is capable of resolving the prominent Pa$\beta$ and Pa$\gamma$ lines in low-$z$ AGNs ($z\lesssim0.5$). Moreover, advancements in fringe-tracking (FT) methodologies (such as the hierarchical FT architecture; see \citealt{Petrov2024}) might enhance the FT limiting magnitude and thus further expand the accessible AGN sample size.
Taken together, we would be able to observe sufficient AGN targets for a statistically relevant combination of high-quality SA with RM. The high-quality SA would also allow to eliminate or better model those targets potentially exhibiting unusual features like PDS~456.
Although our currently estimated Hubble constant still carries a large uncertainty, a sizable RM campaign coordinated with near-infrared interferometry will enable the SARM approach to achieve a competitive precision for the Hubble constant. In light of the geometric nature of the measured distance, the SARM approach will ultimately offer a promising cosmic probe, independent of all other cosmic tools.

\section*{Acknowledgements}
We thank Misty Bentz for kindly sharing the RM light-curve data of NGC 3783 and  Hannah \"{U}bler and the anonymous referee for useful comments on the manuscript. YRL acknowledges financial support from the National Key R\&D Program of China (2021YFA1600404, 2023YFA1607904), the National Natural Science Foundation of China (12273041), the China-Chile Joint Research Fund (CCJRF2310), and from the Youth Innovation Promotion Association CAS. JS is supported by the Fundamental Research Funds for the Central Universities, Peking University (7100604896) and the China Manned Space Program (CMS-CSST-2025-A09). JMW acknowledges financial support from the National Natural Science Foundation of China (11991050, 12333003). LCH is supported by the National Key R\&D Program of China (2022YFF0503401), the National Science Foundation of China (11991052, 12233001), and the China Manned Space Project (CMS-CSST-2021-A04, CMS-CSST-2021-A06). P.D. is supported by the National Key R\&D Program of China (2023YFA1607903) and the China Manned Space Project (CMS-CSST-2025-A07). R.G.P. acknowledges the support of the French Agence Nationale de la Recherche (ANR) through the grant ``AGN  MELBa''  ANR-21-CE31-0011.

We thank the support of the staff of the Lijiang 2.4 m telescope and of the the Centro Astron\'omico Hispanoen Andaluc\'ia (CAHA) 2.2m telescope. This work is based on observations collected at CAHA at Calar Alto, operated jointly by the Andalusian Universities and the Instituto de Astrof\'isica de Andaluc\'ia (CSIC).
This work has made use of the NASA/IPAC Extragalactic Database (NED), which is operated by the Jet Propulsion Laboratory, California Institute of Technology, under contract with the National Aeronautics and Space Administration.

\software{\textsc{Astropy} (\citealt{Astropy2018}), \textsc{BRAINS} (\citealt{brains}), \textsc{CDNest} (\citealt{cdnest}),  PyCALI (\citealt{pycali})}

\appendix
\section{Calculating the BLR Sizes}\label{app_size}
From the transfer function $\Psi_r(\tau)$ in Equation~(\ref{eqn_psir}), the BLR size from the centroid time delay is calculated as
\begin{equation}
R_\tau= c\tau_{\rm cent} = \frac{\int c\tau \Psi_r(\tau)d\tau}{\int \Psi_r(\tau)d\tau},
\label{eqn_rtau}
\end{equation}
which can be recast into
\begin{equation}
R_\tau = \frac{\sum_i c\tau_i\eta_i w_i}{\sum_i \eta_i w_i}.
\end{equation}
The responsivity-weighted size is calculated by summing up all clouds' radii with weights as (\citealt{LW2024})
\begin{equation}
R_{\rm rw} = \frac{\sum_i r_i \eta_i w_i}{\sum_i \eta_i w_i}.
\end{equation}
Similarly,  the emissivity-weighted size is obtained by
\begin{equation}
R_{\rm ew} =\frac{\sum_i r_i w_i}{\sum_i w_i}=R_{\rm BLR},
\end{equation}
which is indeed equal to the $R_{\rm BLR}$ parameter because the weights do not depend on radius. It is easy to show that $R_{\rm rw}$ will be exactly equal to $R_{\rm ew}$ only if $\eta$ is a global constant.

\section{The Full List of Model Parameters and Their Best Estimates}\label{app_parameters}
In Table~\ref{table_full_para}, we list the full SARM analysis parameters and their priors.
In Table~\ref{table_para}, we summarize the best estimates and uncertainties of major parameters, which are assigned by the medians and
68.3\% confidence intervals of the posterior samples, respectively.

\section{Full Fits to Spectroastrometric Data}\label{app_sa}
\begin{figure*}
\centering
\includegraphics[width=0.7\linewidth]{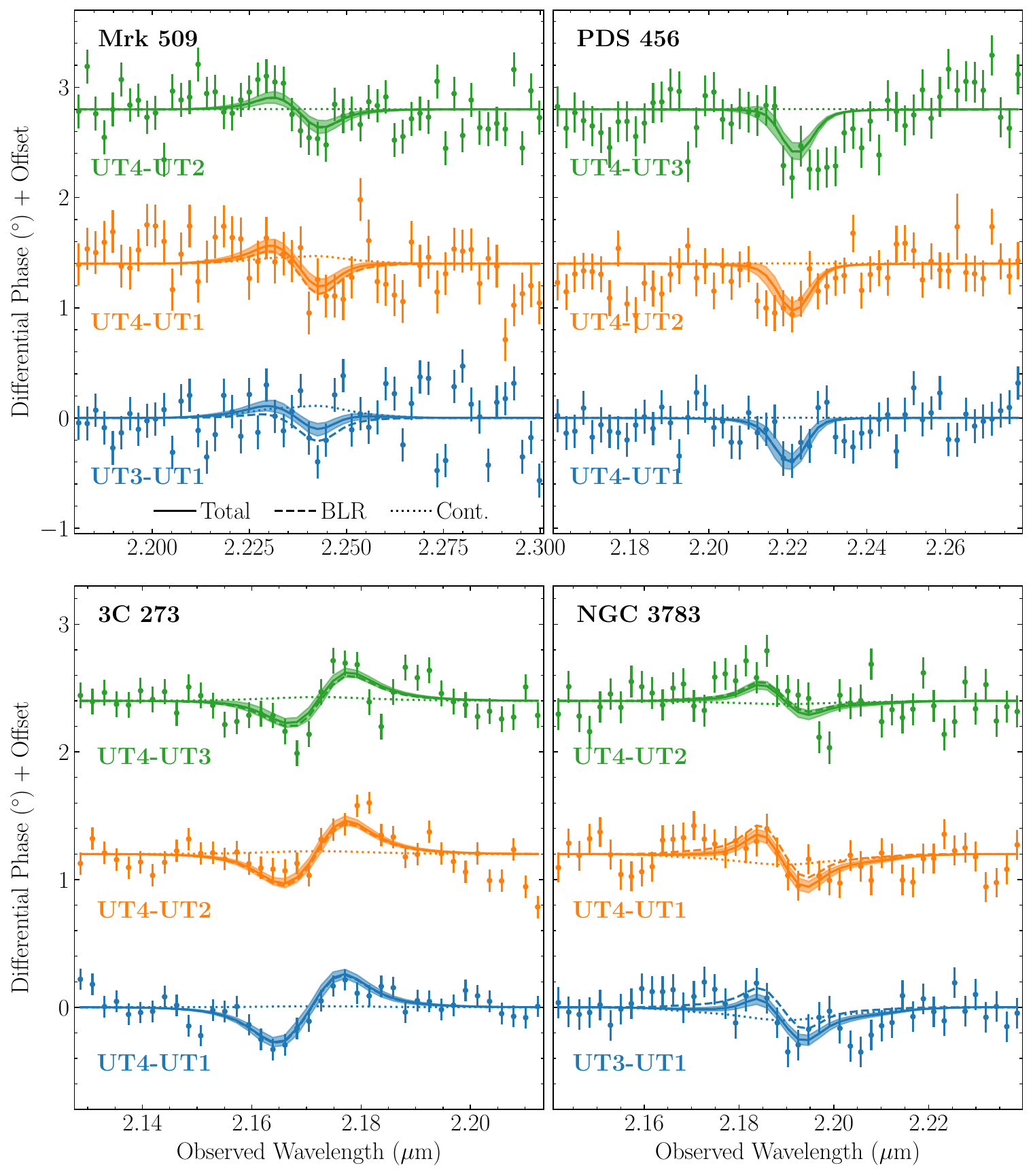}
\caption{Fits to the time-averaged differential phase data (points) of the three baselines showing the strongest BLR signal (dashed lines). The best-fit residual continuum phases are shown with dotted lines. The solid lines with shaded areas represent the best-fits to the total differential phases (BLR+residual continuum) and the uncertainties. The text around each differential phase curve denotes the corresponding baseline name.}
\label{fig_phase}
\end{figure*}

In Figure~\ref{fig_phase}, we show the best fits to the differential phases of the three baselines with the strongest BLR signals for the four AGNs. The differential phases are averaged over epochs.

\section{The Optical Spectra}\label{app_spec}
\begin{figure}[tp!]
\centering
\includegraphics[width=1.0\linewidth]{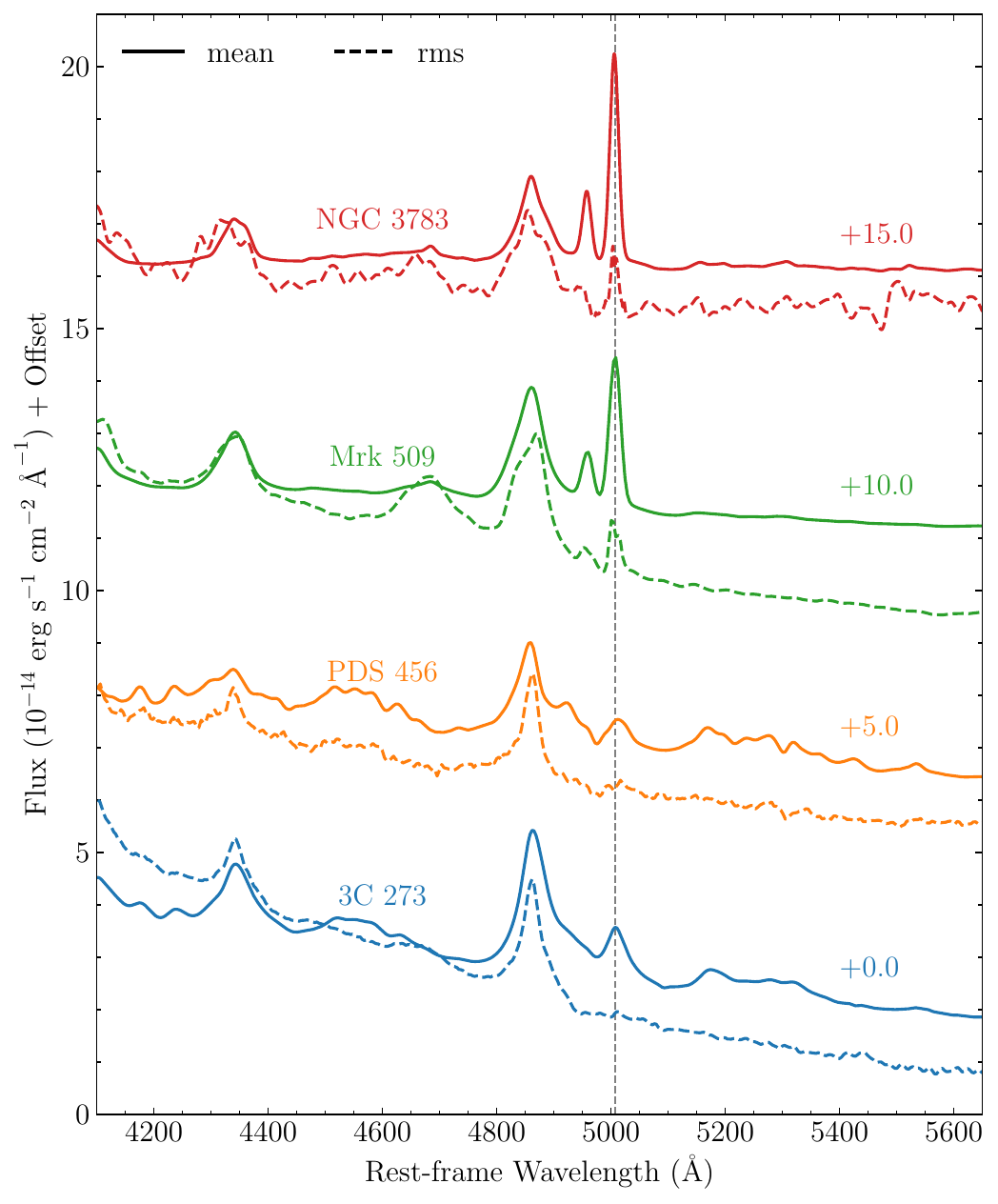}
\caption{The optical mean and rms spectra of the four AGNs from previous RM observations. For each quasar, the solid and dashed lines represent the mean and rms spectra, respectively. The rms spectrum is scaled and shifted to be comparable in flux with the mean spectrum. The data of Mrk 509 and PDS~456 are from \cite{Li2024}, the data of 3C~273 are from \cite{Zhang2019}, and the data of NGC~3783 are from \cite{Stripe1994}. The vertical dashed line represents the wavelength of [\ion{O}{3}]$\lambda$5007. For PDS~456, the small bump around 5007~{\AA} is mainly due to the \ion{Fe}{2} emissions rather than [\ion{O}{3}]$\lambda$5007. }
\label{fig_spec}
\end{figure}

For the sake of completeness, we show the optical mean and rms spectra of Mrk~509, PDS~456, 3C~273, and NGC~3783 from previous RM observations in Figure~\ref{fig_spec}. As observed, PDS~456 and 3C~273 show weak [\ion{O}{3}] and strong \ion{Fe}{2} emissions, which are typical features of super-Eddington AGNs (e.g., \citealt{Boroson1992, Collin2004, Du2014}).

\section{The Upper Limit of the Responsivity}\label{app_etamax}
\begin{figure}[tp!]
\centering
\includegraphics[width=1.0\linewidth]{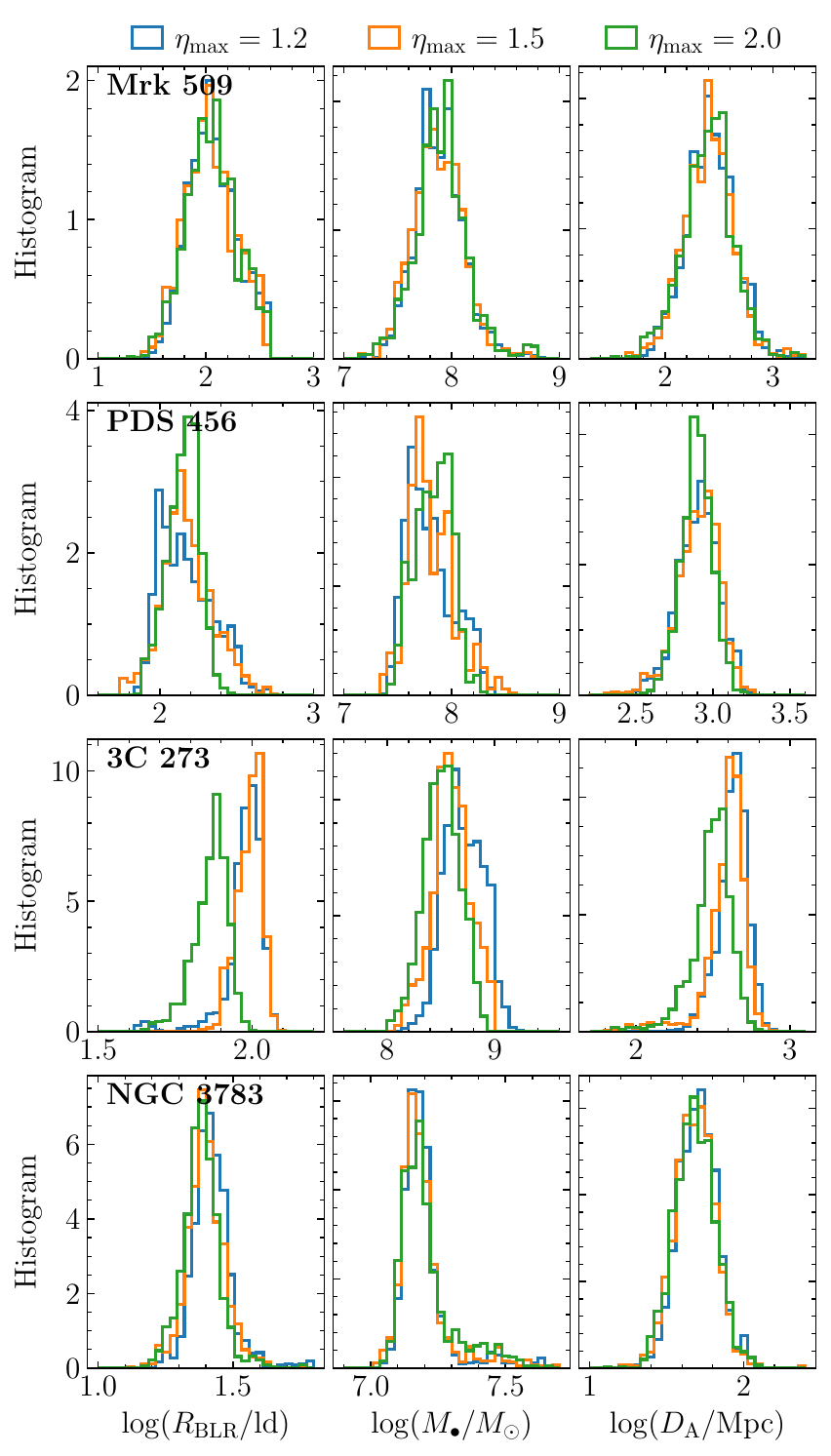}
\caption{A comparison of (left) the obtained BLR size, (middle) the SMBH mass, and (right) the angular-diameter distance for different upper limits of the responsivity ($\eta_{\rm max}$).}
\label{fig_eta_comp}
\end{figure}

In our calculations, the responsivity is parameterized as a power law of radial distance (see Equation~\ref{eqn_eta}). The maximum responsivity is limited to $\eta_{\rm max}=1.5$ according to previous photoionization calculations (e.g., \citealt{Korista2004, Goad2014, Goad2015, Zhang2021, Li2024}). To test whether this limit affects the parameter inference, we rerun the SARM analysis using an upper limit of $\eta_{\rm max}=1.2$ and $2$. In Figure~\ref{fig_eta_comp}, we show a comparison of the obtained BLR size, SMBH mass, and angular-diameter distance for these different upper limits. We find that for Mrk 509, PDS 456, and NGC~3783, the results are quite insensitive to the choice of $\eta_{\rm max}$. For 3C 273, the upper limit of $\eta_{\rm max}=2$ tends to yield a smaller BLR size and thereby a smaller angular-diameter distance. Nevertheless, the differences from those of $\eta_{\rm max}=1.2$ and $1.5$ are about 0.1 dex, still within 1$\sigma$ uncertainties, implying that the influence of $\eta_{\rm max}$ on the distance measurement is not significant. A BLR dynamical model with an inclusion of photoionization models will relax the need for the present simple parameterization for the responsivity and help further improve SARM analysis. This is beyond the scope of this paper and will be deferred to a future work.

\section{Comparison with Time Lags from the Cross-correlation Functions}\label{app_rm}
Given a BLR model, we can directly calculate the centroid time lag of the responsivity-weighted transfer
function using Equation~(\ref{eqn_rtau}). In practice, the transfer function usually has a long tail, arising from the responses of BLR clouds at large distances. This long tail causes the centroid time lag of the transfer function to be longer than that from the traditional cross-correlation function (CCF) of the light curves. The latter is conventionally derived by the centroid of the CCF above 80\% of the peak value (e.g., \citealt{Gaskell1987, Peterson2004}). Therefore, to ensure a meaningful comparison, we calculate the time lag from BLR modeling as follows. We firstly generate mock datasets based on the reconstructed light curves from the BLR model by adopting the same measurement errors and sampling cadences as the observed data. We then use the same CCF method to calculate the time lags of these mock datasets.

\begin{figure}[tp!]
\centering
\includegraphics[width=0.3\textwidth]{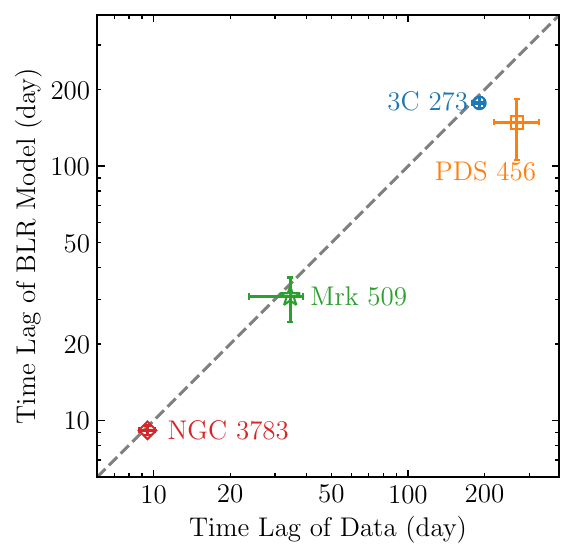}
\caption{A comparison between the centroid time lags derived from BLR dynamical modeling and the traditional CCF analysis of the RM data.}
\label{fig_ccf_comp}
\end{figure}

Figure~\ref{fig_ccf_comp} shows the comparison. The time lags from the two approaches are consistent, except for PDS 456, which shows a relatively shorter time lag from BLR
modeling (but still consistent at $\sim2\sigma$ confidence level). We ascribe this discrepancy to the influence of fine variation patterns in the H$\beta$ light curve of PDS 456 (see Figure~\ref{fig_PDS456}), which appear to lack counterparts in the continuum variations and thereby cannot be reproduced by RM modeling.

%=============================================================================================
%=============================================================================================

\begin{deluxetable*}{lccl}
\tablecaption{The Full List of Model Parameters and Priors.\label{table_full_para}}
\tablewidth{0.95\textwidth}
\tablehead{
\colhead{~~~Parameter~~~~~~~~~}  & \colhead{~~~~~~~~~~~~~~Prior~~~~~~~~~~} & \colhead{~~~~~~~~~~Range~~~~~~~~~~~~~~} & \colhead{Note}
}
\startdata
$R_{\rm BLR}$ (ld)                &  LogUniform & (1, $\Delta T$/2) &  Emissivity-weighted mean BLR size\\
$F$                                      &  Uniform    & (0, 1)               &  Fraction of the inner BLR edge in units of $R_{\rm BLR}$\\
$\beta$                                  &  Uniform    & (0, 2)               &  Shape parameter of radial distribution of BLR clouds\\
$\theta_{\rm inc}$ ($^\circ$)            &  CosUniform & (0, 90)              &  Inclination angle\\
$\theta_{\rm opn}$ ($^\circ$)            &  Uniform    & (0, 90)              &  Openning angle\\
$\kappa$                                 &  Uniform    & (-1/2, 1/2)          &  Azimuthal anisotropic parameter of cloud emission\\
$\gamma$                                 &  Uniform    & (0, 5)               &  Clustring parameter of BLR clouds towards the surfaces\\
$\xi$                                    &  Uniform    & (0, 1)               &  Transparency of the BLR equatorial plane\\
$\eta_0$                                 &  Uniform    & (0, 1.5)             &  Constant coefficient of responsivity \\
$\eta_1$                                 &  Uniform    & (0, 1.5)             &  Power-law amplitude of  responsivity\\
$\alpha$                                 &  Uniform    & (0, 5)               &  Power-law index of  responsivity\\
PA ($^\circ$ E of N)                     &  Uniform    & (0, 360)             &  Position angle\\
$x_c$ (ld)                        &  Uniform    & (0, 100)             &  Longitudinal offset of the continuum emission projected on the sky\\
$y_c$ (ld)                        &  Uniform    & (0, 100)             &  Latitudinal offset of the continuum emission projected on the sky\\
$D_{\rm A}$ (Mpc)                        &  LogUniform & (10, 10$^4$)         &  Angular-diameter distance\\
$M_\bullet$ ($M_\odot$)                  &  LogUniform & (10$^6$, 10$^9$)     &  Black hole mass\\
$f_{\rm ellip}$                          &  Uniform    & (0, 1)               &  Fraction of clouds in elliptical orbits\\
$f_{\rm flow}$                           &  Uniform    & (0, 1)               &  Flag for determining inflowing or outflow orbits\\
$\theta_{e}$ ($^\circ$)                  &  Uniform    & (0, 90)              &  Rotation angle of inflow or outflow orbits\\
$\bar F_c$                               &  Uniform    & (0, 2)               &  Constant component of the continuum light curve\\
%$\sigma_{\rm d}$                         &  LogUniform & (10$^{-3}$, 0.1)     &  Long-term standard deviation of DRW variation \\
%$\tau_{\rm d}$                           &  LogUniform & (1, $\Delta T$)      &  Typical damping timescale of DRW variation\\
%$\bm{u}_q$                               &  Gaussian   & \nodata              &  Deviations of long-term trend of the continuum light curve\\
%$\bm{u}_s$                               &  Gaussian   & \nodata              &  Series of deviations of the continuum light curve
\enddata
\tablecomments{$\Delta T$ denotes the time span of the emission line light curve. ``LogUniform'' denotes a uniform prior distribution for the logarithm of the parameter, and ``CosUniform'' denotes a uniform prior distribution for the cosine of the parameter.}
\end{deluxetable*}

\begin{deluxetable*}{ccccc}
\tablecaption{Inferred Parameter Values and Uncertainties.\label{table_para}}
\tablehead{
\colhead{Parameter}  & \colhead{Mrk 509} & \colhead{PDS 456\tablenotemark{\footnotesize a}}  & \colhead{3C 273}  & \colhead{NGC 3783}
}
\startdata
$\log(R_{\rm BLR}/\rm ld)$     & $2.03_{-0.22}^{+0.28}$& $2.16_{-0.14}^{+0.19}$& $1.93_{-0.04}^{+0.05}$& $1.40_{-0.06}^{+0.07}$\\
$F$                            & $0.17_{-0.07}^{+0.07}$& $0.05_{-0.01}^{+0.01}$& $0.16_{-0.03}^{+0.03}$& $0.24_{-0.02}^{+0.03}$\\
$\beta$                        & $1.21_{-0.26}^{+0.31}$& $1.83_{-0.09}^{+0.09}$& $1.16_{-0.10}^{+0.11}$& $1.96_{-0.06}^{+0.03}$\\
$\theta_{\rm inc}~(^\circ)$    & $44_{-15}^{+25}$& $47_{-13}^{+36}$& $8_{-1}^{+2}$& $67_{-15}^{+5}$\\
$\theta_{\rm opn}~(^\circ)$    & $36_{-14}^{+32}$& $58_{-13}^{+28}$& $13_{-4}^{+4}$& $70_{-13}^{+4}$\\
$\kappa$                       & $-0.29_{-0.15}^{+0.32}$& $-0.37_{-0.10}^{+0.09}$& $0.19_{-0.39}^{+0.23}$& $-0.16_{-0.12}^{+0.14}$\\
$\gamma$                       & $2.73_{-1.24}^{+1.49}$& $3.64_{-1.73}^{+1.01}$& $2.48_{-1.17}^{+1.73}$& $3.34_{-1.26}^{+1.08}$\\
$\xi$                          & $0.29_{-0.21}^{+0.31}$& $0.13_{-0.08}^{+0.18}$& $0.13_{-0.10}^{+0.24}$& $0.82_{-0.20}^{+0.13}$\\
$\eta_0$                       & $0.99_{-0.41}^{+0.35}$& $0.34_{-0.18}^{+0.24}$& $0.00_{-0.00}^{+0.00}$& $0.95_{-0.20}^{+0.16}$\\
$\eta_1$                       & $0.76_{-0.49}^{+0.50}$& $1.22_{-0.40}^{+0.19}$& $0.43_{-0.18}^{+0.37}$& $0.45_{-0.35}^{+0.67}$\\
$\alpha$                       & $2.23_{-1.67}^{+1.71}$& $2.81_{-1.69}^{+1.61}$& $4.85_{-0.21}^{+0.11}$& $2.45_{-1.66}^{+1.73}$\\
PA ($^\circ$ E of N)\tablenotemark{\footnotesize b}           & $196_{-31}^{+29}$& $305_{-7}^{+5}$& $389_{-49}^{+23}$& $134_{-16}^{+16}$\\
$x_c$ (ld)                     & $2.64_{-15.32}^{+19.76}$& 0 & $-2.44_{-2.76}^{+2.75}$& $0.43_{-2.03}^{+2.56}$\\
$y_c$ (ld)                     & $-33.27_{-32.64}^{+22.07}$& 0 & $3.91_{-4.28}^{+4.50}$& $4.95_{-2.56}^{+3.11}$\\
$\log(D_{\rm A}/\rm Mpc)$      & $2.41_{-0.23}^{+0.22}$& $2.91_{-0.13}^{+0.13}$& $2.58_{-0.11}^{+0.09}$& $1.68_{-0.13}^{+0.12}$\\
$\log(M_\bullet/M_\odot)$      & $7.86_{-0.23}^{+0.24}$& $7.76_{-0.14}^{+0.23}$& $8.55_{-0.15}^{+0.21}$& $7.17_{-0.05}^{+0.07}$\\
$f_{\rm ellip}$                & $0.32_{-0.22}^{+0.21}$& $0.18_{-0.12}^{+0.21}$& $0.62_{-0.15}^{+0.14}$& $0.57_{-0.14}^{+0.10}$\\
$f_{\rm flow}$                 & $0.70_{-0.20}^{+0.20}$& $0.57_{-0.37}^{+0.31}$& $0.54_{-0.38}^{+0.32}$& $0.64_{-0.42}^{+0.25}$\\
$\theta_{\rm e}~(^\circ)$      & $32_{-20}^{+24}$& $10_{-7}^{+18}$& $31_{-22}^{+23}$& $11_{-7}^{+17}$\\
\enddata
\tablenotetext{\footnotesize a}{The continuum phase offset of PDS~456 is set to zero. }
\tablenotetext{\footnotesize b}{Note that the position angle (PA) has a 180$^\circ$ ambiguity. }
\end{deluxetable*}

\end{document}